\documentclass{aa}  
\usepackage{graphicx}
\usepackage{float}
\usepackage{indentfirst}
\usepackage{lscape}
\usepackage{longtable}
\usepackage{amsmath}
\usepackage{amssymb}
\usepackage{graphicx}
\usepackage{lastpage}
\usepackage{natbib}
\usepackage{txfonts}
\begin{document}
%
   \title{Evolution of Chemistry in the envelope of HOt CorinoS (ECHOS)}
   
      \subtitle{II. The puzzling chemistry of isomers as revealed by the HNCS/HSCN ratio}

   \author{G. Esplugues
          \inst{1},   
          M. Rodríguez-Baras
          \inst{1},      
          D. Navarro-Almaida
          \inst{2},  
          A. Fuente  
          \inst{3},
          P. Fernández-Ruiz
          \inst{3},
          S. Spezzano 
          \inst{4},
          M. N. Drozdovskaya 
          \inst{5},
          Á. Sánchez-Monge 
          \inst{6, 7},
          P. Caselli 
          \inst{4},
          P. Rivière-Marichalar  
          \inst{1}, and
          L. Beitia-Antero
          \inst{8, 9}
         }

   \institute{Observatorio Astron\'omico Nacional (OAN), Alfonso XII, 3, 28014. Madrid. Spain\\
              \email{g.esplugues@oan.es}
        \and
Universit\'{e} Paris-Saclay, Universit\'e Paris Cit\'{e}, CEA, CNRS, AIM, F-91191 Gif-sur-Yvette, France       
        \and
Centro de Astrobiolog\'ia (CAB), INTA-CSIC, Carretera de Ajalvir Km. 4, Torrej\'on de Ardoz, 28850, Madrid, Spain 
		\and
Max-Planck-Institut für extraterrestrische Physik, 85748 Garching, Germany
		\and
Physikalish-Meteorologisches Observatorium Davos und Weltstrahlungszentrum (PMOD/WRC), Dorfstrasse 33, CH-7260, Davos Dorf, Switzerland   	
		\and
Institut de Ciències de l'Espai (ICE, CSIC), Can Magrans s/n, E-08193, Bellaterra, Barcelona, Spain
		\and
Institut d'Estudis Espacials de Catalunya (IEEC), Barcelona, Spain
		\and
Departamento de Estad\'istica e Investigaci\'on Operativa, Facultad de Ciencias Matem\'aticas, Universidad Complutense de Madrid, Spain
		\and
Joint Center for Ultraviolet Astronomy, Universidad Complutense de Madrid, Avda Puerta de Hierro s/n, 28040, Madrid, Spain	
}


 
  \abstract   
   {The observational detection of some metastable isomers in the interstellar medium with abundances comparable to those of the most stable isomer, or even when the stable isomer is not detected, highlights the importance of non-equilibrium chemistry. This challenges our understanding of the interstellar chemistry and shows the need to study isomeric forms of molecular species to constrain chemical processes occurring in the interstellar medium.    
   } 
   {Our goal is to study the chemistry of isomers through the sulphur isomer pair HNCS and HSCN, since HSCN has been observed in regions where its stable isomer has not been detected, and the observed HNCS/HSCN ratio seems to significantly vary from cold to warm regions.   
   }
   {We used the Nautilus time-dependent gas-grain chemical code to model the formation and destruction paths of HNCS and HSCN in different astrochemical scenarios, as well as the time evolution of the HNCS/HSCN ratio. We also analysed the influence of the environmental conditions on their chemical abundances. 
   }
   {We present an observational detection of the metastable isomer HSCN in the Class I object B1-a ($N$=(1.1$\pm$0.6)$\times$10$^{12}$ cm$^{-2}$), but not of the stable isomer HNCS, despite HNCS lying 3200 K lower in energy than HSCN. Our theoretical results show an HNCS/HSCN ratio sensitive to the gas temperature and the evolutionary time, with the highest values obtained at early stages ($t$$\lesssim$10$^4$ yr) and low ($T_{\mathrm{g}}$$\lesssim$20 K) temperatures. A more detailed analysis also shows that the main mechanism forming HNCS in young ($t$$<$10$^5$ yr) and cold ($T_{\mathrm{g}}$$=$10 K) objects at moderate-low densities ($n$$_{\mathrm{H}}$$\leq$10$^5$ cm$^{-3}$) is a grain surface reaction (the chemical desorption reaction N$_{\mathrm{solid}}$+HCS$_{\mathrm{solid}}$$\rightarrow$HNCS), unlike previous predictions that suggest only gas-phase chemistry for its formation. However, for the formation of its metastable isomer HSCN, over the same time range and physical conditions, we find that both surface and gas-phase reactions (through the ion H$_2$NCS$^+$) play important roles. In warmer ($T_{\mathrm{g}}$$\geq$50 K) regions, the formation of this isomer pair is only dominated by gas-phase chemistry through the ions H$_2$NCS$^+$ (mainly at low densities) and HNCSH$^+$ (mainly at high densities).  
}   
   {The results  suggest a different efficiency of the isomerisation processes depending on the source temperature. The progressive decrease of HNCS/HSCN with gas temperature at early evolutionary times derived from our theoretical results indicates that this ratio may be used as a tracer of cold young objects. This work also demonstrates the key role of grain surface chemistry in the formation of the isomer pair HNCS and HSCN in cold regions, as well as the importance of the ions H$_2$NCS$^+$ and HNCSH$^+$ in warm/hot regions. Since most of the interstellar regions where HSCN is detected are cold regions (starless cores, Class 0/I objects), a larger sample including sources characterised by high temperatures (e.g. hot cores) are needed to corroborate the theoretical results.     
}

   \titlerunning{A theoretical study of HNCS/HSCN}
   \authorrunning{G. Esplugues et al.}
   \maketitle
%

\section{\textbf{Introduction}}

The interstellar medium (ISM) presents a large variety of physical conditions with temperatures ranging between $\sim$10-10$^4$ K, densities varying from $\sim$10–10$^{8}$ cm$^{-3}$, and energetic radiation (in some cases even ionising radiation) \citep[]{Yamamoto2017}. Thermodynamic equilibrium requires all species to collide with each other frequently enough. The collisional timescale is determined \citep[e.g.][]{Klein2008} by

\begin{equation}
\tau_{\mathrm{c}} \approx  \displaystyle{ {\frac{\lambda}{v}}} \approx 1.3\times10^{11} \left(\displaystyle{ \frac{T}{{\mathrm{K}}}}\right)^{-1/2} \left(\displaystyle{ \frac{n_{\mathrm{H}}}{1 {\mathrm{cm}}^{-3}}}\right)^{-1} {\mathrm{s}}, 
\end{equation}

\noindent where $\lambda$ is the mean free path of the particle, $v$ is the typical particle velocity, $T$ is the temperature, and $n$$_{\mathrm{H}}$ is the atomic H density. Adopting $T$=100 K and $n$$_{\mathrm{H}}$= 1 cm$^{-3}$, we obtain about {\bf{one}} collision in $\sim$500 years. Therefore, the ISM can be considered far from being in thermodynamic equilibrium. In spite of this, we find very large and complex chemical species in the ISM, in particular in dense clouds, some of them having even a prebiotic importance. This is the case of many detected molecules, such as the simplest sugar glycolaldehyde \citep[]{Hollis2000}, urea \citep[]{Belloche2019}, which is a possible precursor of ribonucleotides \citep[]{Becker2019}, and amino acetonitrile, vinyl amine, and ethyl amine, which are precursors of amino acids \citep[]{Belloche2008, Zeng2021}. 

Another proof of non-equilibrium in the ISM is the detection of metastable isomers. Of all interstellar molecules that could have isomers ({\bf{i.e.}} molecules containing three or more atoms), about 30$\%$ have detected isomeric counterparts \citep[]{Hollis2005}, which implies a very high presence of isomerism in the ISM. 
In literature, there are several detections of species with much higher energies than their stable isomers. This is, for instance, the case for HNC, whose energy is 7400 K higher than that of HCN \citep[]{Adande2010}, but it is still observed in cold molecular clouds (where the kinetic temperature is well below 100 K) with nearly the same abundance as HCN \citep[]{Hirota1998}. In warm molecular clouds, however, the HNC/HCN abundance ratio is significantly lower than unity \citep[]{Irvine1984, Churchwell1984, Schilke1992, Hirota1998} because HNC abundance decreases with the kinetic temperature \citep[]{Marcelino2010}.

A similar scenario holds for the isomer pair HNCS and HSCN. Isothiocyanate, HNCS (isothiocyanic acid), is a linear molecule that was first detected by \cite{Frerking1979} in Sgr B2(OH). Its metastable isomer, HSCN (thiocyanic acid), was also first detected in the Sagittarius region, in particular in Sgr B2(N), by \cite{Halfen2009}. HSCN lies over 3200 K (about 6 kcal mol$^{-1}$) higher in energy than HNCS and has a bent structure \citep[]{Wierzejewska2003, Adande2010}. 
The HNCS/HSCN ratio is found to be $\sim$2-7 in Sgr\,B2 with the lowest values near the warm hot cores \citep[]{Adande2010}. In particular, in Sgr\,B2, HNCS and HSCN present extended spatial distributions over a 6$'$$\times$3$'$ region \citep[]{Adande2010}, where the gas temperature does not exceed 50 K \citep[]{deVicente1997}. These observations show the largest HNCS/HSCN ratios at the 2N position, where the gas is relatively quiescent.
These results are puzzling since they suggest a different efficiency of the isomerisation processes depending on the temperature of the clouds. 

Altogether this shows the important role of non-equilibrium chemistry in the interstellar medium, as well as the study of isomerism, since the analysis of their formation and destruction chemical processes may provide clues to better understanding the complexity of the ISM. 

In this work, we carried out a comprehensive study of the sulphur isomer pair HNCS and HSCN to analyse the influence of the environmental conditions on their abundances. We present the chemical model that we have used in Sect. \ref{section:model}. In Sect. \ref{section:model_results}, we present the model results concerning the evolution of the HNCS and HSCN abundances and of their ratio. A discussion, including an exhaustive analysis of the main mechanisms forming and destroying this isomer pair, as well as a comparison with other isomeric ratios, is shown in Sect. \ref{Discussion}. We also present (Sect. \ref{Observations}) observations of HSCN in the Solar-type protostar (Class I) B1-a carried out with the Yebes-40m telescope in this paper. We analyse these observational results by comparing them with those found in other type of sources. We finally summarise our conclusions in Sect. \ref{section:summary}.    

\begin{table}
\caption{Abundances with respect to total hydrogen nuclei considered in the chemical code Nautilus.
}             
\centering 
\begin{tabular}{l l l l l}     
\hline\hline       
Species &  Abundance   & Reference                       \\ 
\hline 
He      & 9.0$\times$10$^{-2}$   &  (1) \\
O       & 2.4$\times$10$^{-4}$   &  (2) \\
Si$^+$  & 8.0$\times$10$^{-9}$   &  (3) \\ 
Fe$^+$  & 3.0$\times$10$^{-9}$   &  (3) \\ 
S$^+$   & 1.5$\times$10$^{-6}$   &  (4)  \\
Na$^+$  & 2.0$\times$10$^{-9}$   &  (3)  \\
Mg$^+$  & 7.0$\times$10$^{-9}$   &  (3)  \\
P$^+$   & 2.0$\times$10$^{-10}$  &  (3)  \\
Cl$^+$  & 1.0$\times$10$^{-9}$   &  (3)   \\
F$^+$   & 6.7$\times$10$^{-9}$   &  (5)  \\
N       & 6.2$\times$10$^{-5}$   &  (6) \\
C$^+$   & 1.7$\times$10$^{-4}$   &  (7) \\
H$_2$  & 0.5                     &  (8) \\
\hline 
\label{table:abundances_Nautilus}                 
\end{tabular}
\tablefoot{
References: (1) Taken from \cite{Asplund2009} and \cite{Wakelam2008}. (2) Taken from \cite{Wakelam2008} and \cite{Navarro-Almaida2021}. (3) As in the low-metal abundance case from \cite{Graedel1982} and \cite{Morton1974}. (4) We considered a depletion factor of 10  with respect to the sulphur cosmic elemental abundance of 1.5×10$^{-5}$ to take into account the recent sulphur depletion results from \cite{Esplugues2022}, \cite{Esplugues2023}, and \cite{Fuente2023}. (5) Taken from \cite{Neufeld2005}. (6) Taken from \cite{Navarro-Almaida2021} and \cite{Jimenez-Serra2018}. (7) Taken from \cite{Navarro-Almaida2021}. (8) Taken from \cite{Wakelam2021}.\\
} 
\\
\end{table}

\section{Chemical model}
\label{section:model}

To theoretically study the chemistry forming and destroying HNCS and HSCN, we used the chemical model Nautilus \citep[]{Ruaud2016}, which is a three-phase model (gas, grain surface, and grain mantle). Nautilus solves the kinetic equations for both the gas phase species and the grain surface species of interstellar dust grains, and computes the time evolution of chemical abundances. In the case of surface species, Nautilus distinguishes among species in the most external layers (in particular, two monolayers), the so-called surface species, and the species below these layers (the so-called mantle species). In particular, the species on the surface becomes a mantle species when two new monolayers of material are accreted above it. These two monolayers become the new surface of the ice.

With respect to chemical processes, the ones occurring in the gas phase that are included in Nautilus are bimolecular reactions (neutral-neutral and ion-neutral reactions); direct cosmic-ray ionization or dissociation; ionisation and dissociation by UV photons; ionisation and dissociation produced by photons induced by cosmic-ray interactions with the medium \citep[]{Prasad1983}; and electronic re-combinations. Regarding chemical-grain surface processes, this version of the code includes the sputtering of grains by cosmic-ray particles (CR sputtering) and non-thermal desorption (distinguishing among three types of desorption: photo-desorption, chemical desorption, and cosmic-ray heating). These three types of desorption can only occur for surface species, while CR sputtering can take place for both surface and mantle species. For more details about the version used in this paper we invite the reader to consult \cite{Wakelam2021}.

The chemical network used in Nautilus is based on the KInetic Database for Astrochemistry (KIDA\footnote{https://kida.astrochem-tools.org/}). In particular, it is the chemical network kida.uva.2022{\bf{;}} this contains the updates presented in \cite{Wakelam2019} and new non-thermal desorption mechanisms, such as cosmic-ray sputtering, as described in \cite{Wakelam2021}. In all models, we adopted the initial abundances shown in Table \ref{table:abundances_Nautilus}.

\section{Model results}
\label{section:model_results}

\begin{figure*}
\centering
\includegraphics[scale=0.32, angle=0]{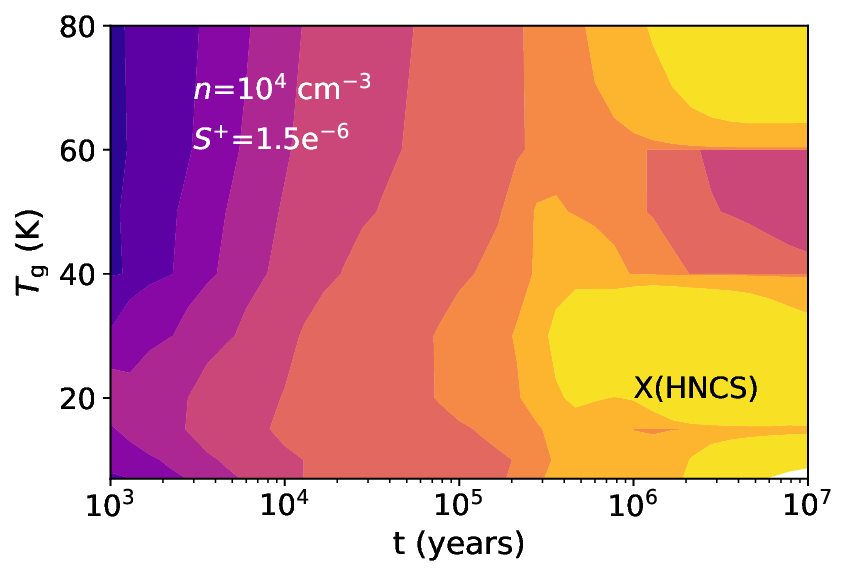}  
\hspace{-0.2cm}
\includegraphics[scale=0.32, angle=0]{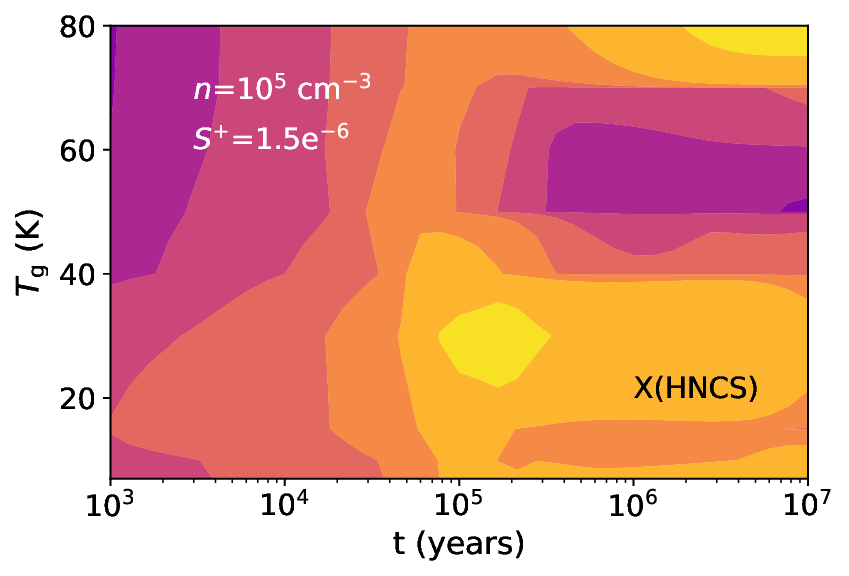}  
\hspace{-0.2cm}
\includegraphics[scale=0.32, angle=0]{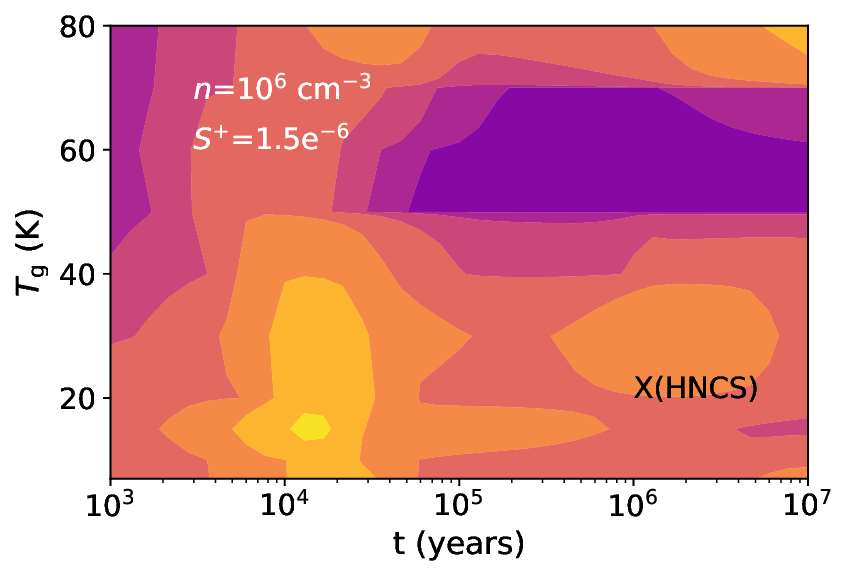} 
\hspace{-0.2cm}
\includegraphics[scale=0.32, angle=0]{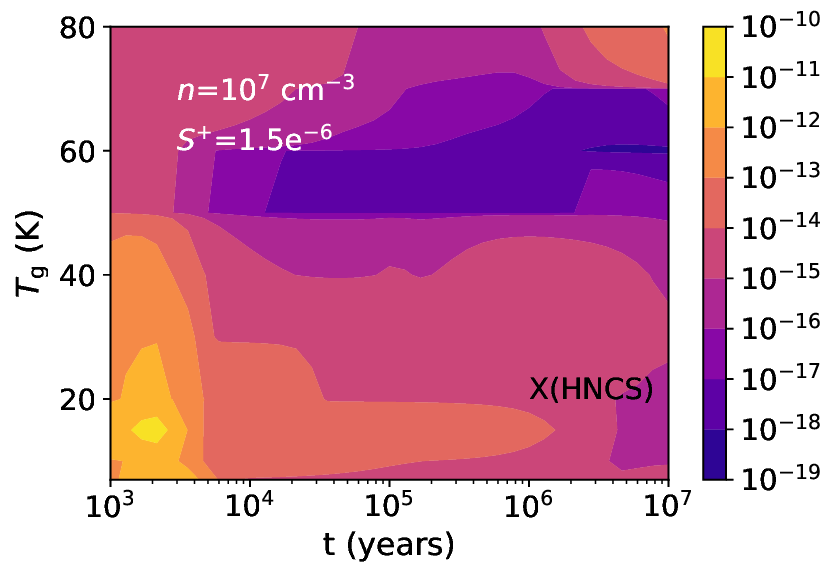}  

\includegraphics[scale=0.32, angle=0]{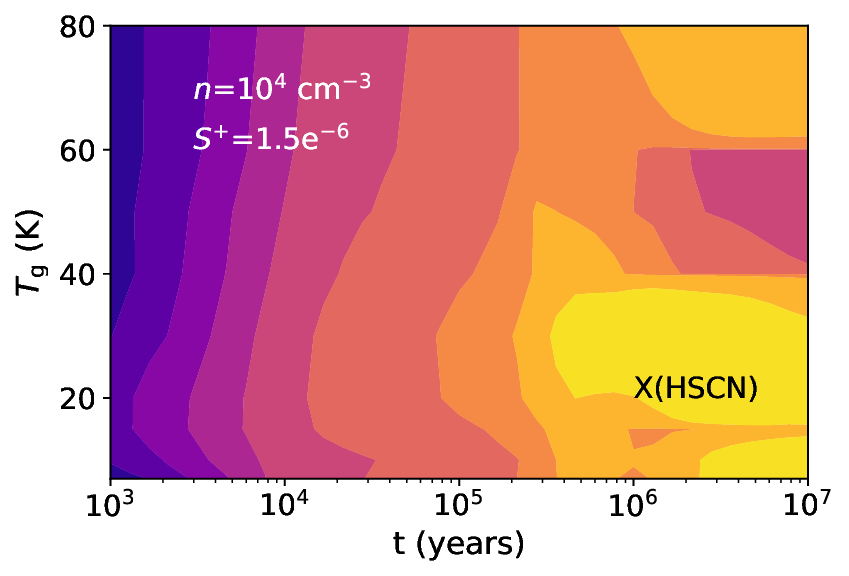}  
\hspace{-0.2cm}
\includegraphics[scale=0.32, angle=0]{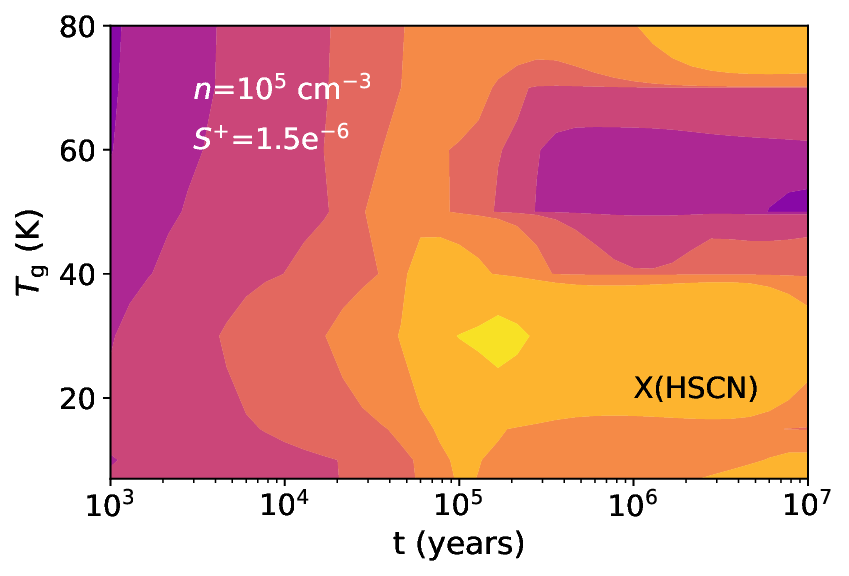}  
\hspace{-0.2cm}
\includegraphics[scale=0.32, angle=0]{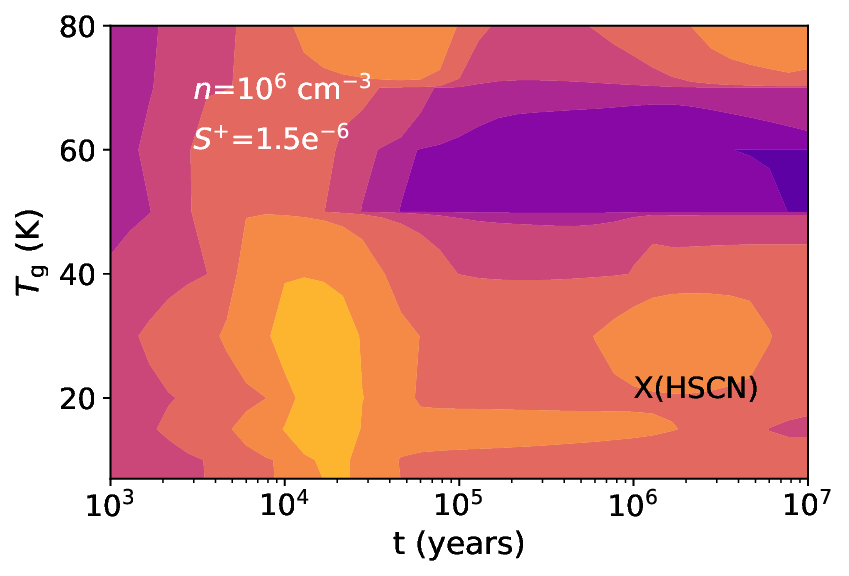}  
\hspace{-0.2cm}
\includegraphics[scale=0.32, angle=0]{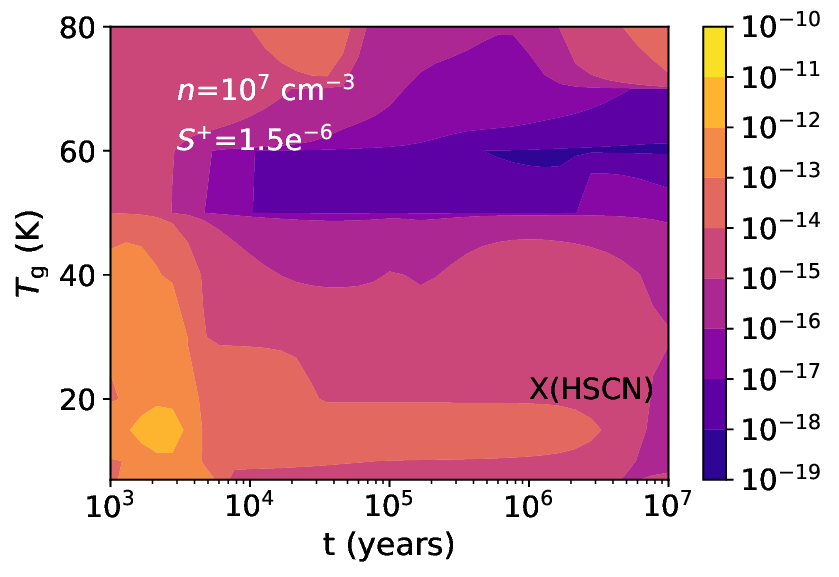}  

\hspace{-0.5cm}
\\
\caption{Evolution of HNCS (top) and HSCN (bottom) fractional abundance as a function of time for an initial sulphur abundance  (relative to $n_{\mathrm{H}}$) $S$$^+$=1.5$\times$10$^{-6}$, a CR ionisation rate $\xi$=1.3$\times$10$^{-17}$ s$^{-1}$, and a hydrogen number density $n_{\mathrm{H}}$=10$^4$, 10$^5$, 10$^6$, and 10$^7$ cm$^{-3}$.
}
\label{figure:abundances}
\end{figure*}

\subsection{HNCS and HSCN abundance evolution}
\label{subsection:abundance}

We computed several Nautilus models in order to study the sensitivity of the HNCS and HSCN abundances to different physical parameters: gas density, gas temperature, and temporal evolution.

In Figure \ref{figure:abundances}, we show the abundances of HNCS and HSCN as a function of time over the course of 10 million years when varying the density from $n_{\mathrm{H}}$=10$^4$ to 10$^7$ cm$^{-3}$ and the gas temperature from $T_{\mathrm{g}}$=7 to 80 K, which are typical conditions of the gas in star-forming dense cores. We first highlight the fact that, for densities of $n_{\mathrm{H}}$$\leq$10$^5$ cm$^{-3}$, both isomers reach the same maximum abundance ($X$$\sim$10$^{-10}$) at the same time. At higher densities, we find that the maximum abundance reached by HNCS is about one order of magnitude higher than the one reached by HSCN. This result suggests that low-density interstellar environments promote the formation of the metastable isomer HSCN.    

Another effect observed for both molecules when varying the density is that the time at which the maximum abundance is reached decreases as density increases. In particular, we obtain the maximun values of HNCS and HSCN at $t$$\sim$5$\times$10$^5$, $\sim$10$^5$, $\sim$10$^4$, and $t$$\sim$10$^3$ yr for $n_{\mathrm{H}}$=10$^4$, 10$^5$, 10$^6$, and 10$^7$ cm$^{-3}$, respectively. In the case of low-density $n_{\mathrm{H}}$=10$^4$ cm$^{-3}$, we also observe that, when the gas temperature is in the range of 10$\lesssim$$T_{\mathrm{g}}$$\lesssim$35 K, the gas-phase abundances of both isomers reach their maximum values and then remain stable from $\sim$5$\times$10$^5$ years on. For the rest of the cases, however, especially the ones with the highest densities, once the maximum abundance is reached for either molecule, it rapidly decreases. This may be due to the lower number of molecular interactions in the case of low density environments, thus allowing the preservation of HNCS and HSCN.

Another physical parameter that we analyse in Fig. \ref{figure:abundances} is the gas temperature. In this case, we observe that the highest values of $X_{\mathrm{HNCS}}$ and $X_{\mathrm{HSCN}}$ are reached for $T_{\mathrm{g}}$$\lesssim$40 K, when $n_{\mathrm{H}}$$\leq$10$^5$ cm$^{-3}$, and for even lower temperatures ($T_{\mathrm{g}}$$\lesssim$20 K) when $n_{\mathrm{H}}$$>$10$^6$ cm$^{-3}$.

In general, as derived from Fig. \ref{figure:abundances}, the highest abundances of the isomers HNCS and HSCN are mostly found at late evolutionary stages ($t$$\gtrsim$10$^5$ yr) for low densities, and at early evolutionary stages ($t$$<$10$^5$ yr) for interstellar regions with high densities, as well as at low/warm temperatures ($T_{\mathrm{g}}$$<$40 K) for most of the cases.

\begin{figure*}
\centering
\includegraphics[scale=0.33, angle=0]{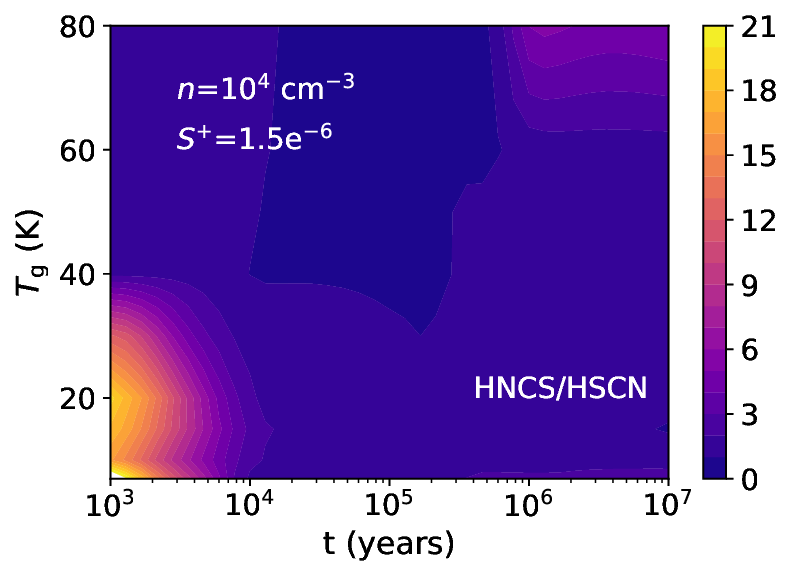}  
\hspace{-0.2cm}
\includegraphics[scale=0.33, angle=0]{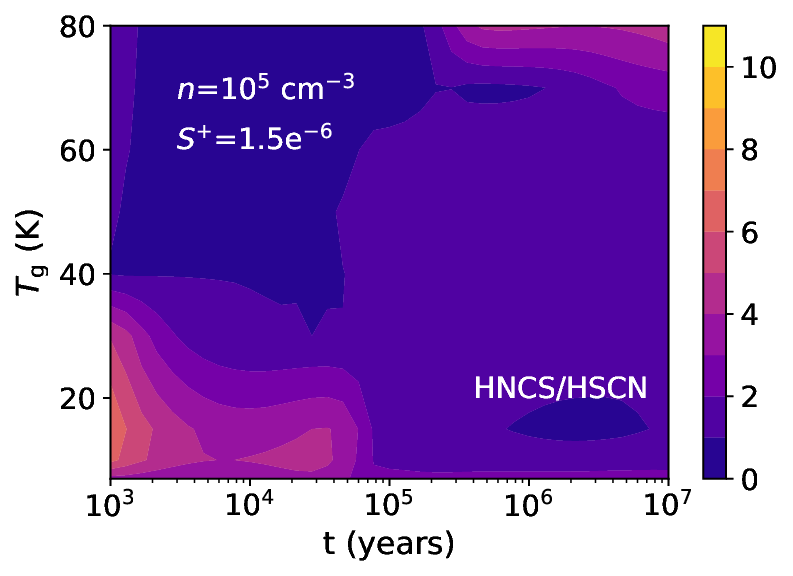}  
\hspace{-0.2cm}
\includegraphics[scale=0.33, angle=0]{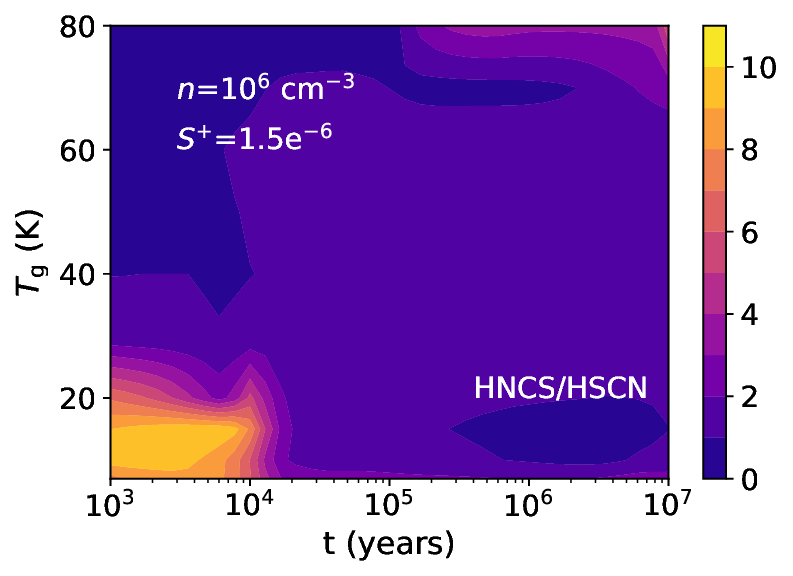} 
\hspace{-0.2cm}
\includegraphics[scale=0.33, angle=0]{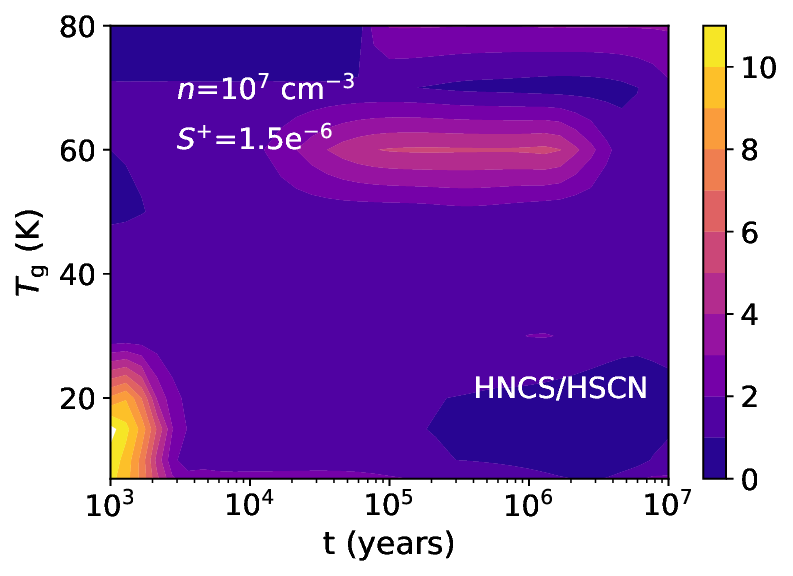}  

\hspace{-0.5cm}
\\
\caption{Evolution of HNCS/HSCN ratio (colour bar) as a function of time and temperature for a CR ionisation rate $\xi$=1.3$\times$10$^{-17}$ s$^{-1}$, an initial sulphur abundance (relative to $n_{\mathrm{H}}$) $S$$^+$=1.5$\times$10$^{-6}$, and a hydrogen number density $n_{\mathrm{H}}$=10$^4$, 10$^5$, 10$^6$, and 10$^7$ cm$^{-3}$.
}
\label{figure:HNCS/HSCN}
\end{figure*}

\begin{figure*}
\centering
\includegraphics[scale=0.32, angle=0]{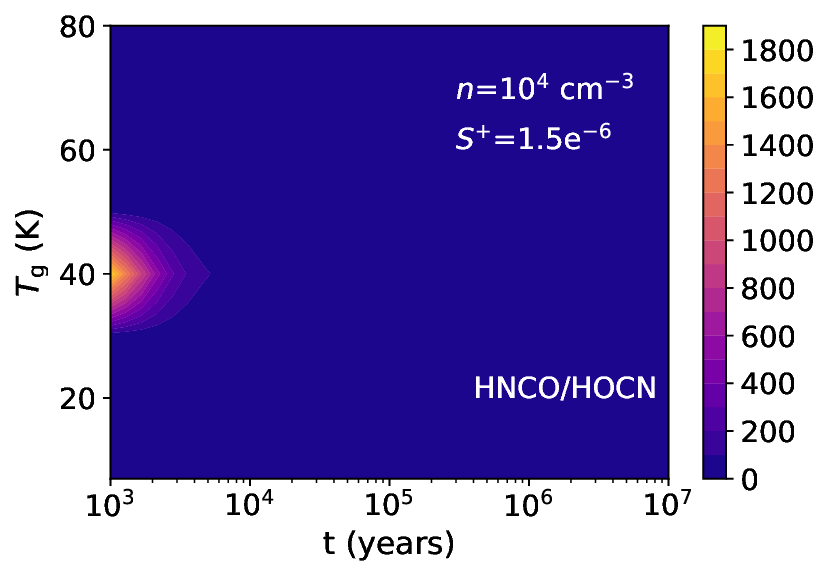}  
\hspace{-0.2cm}
\includegraphics[scale=0.32, angle=0]{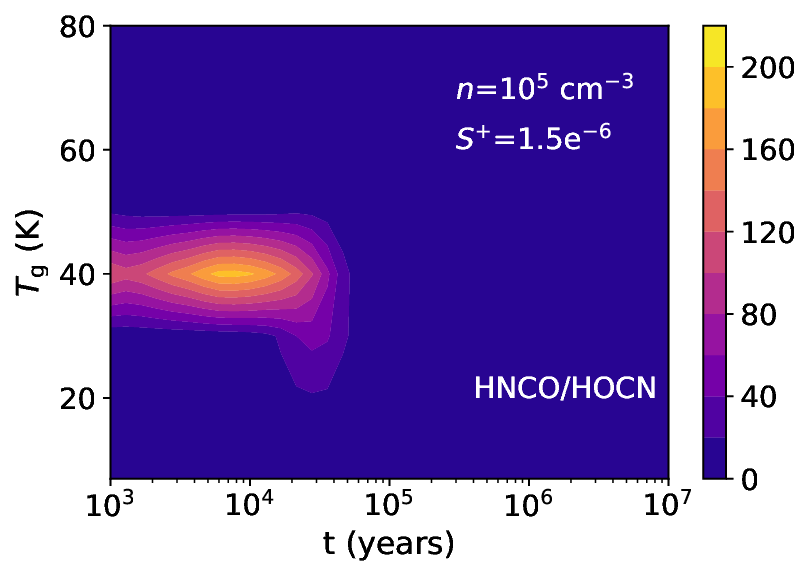}  
\hspace{-0.2cm}
\includegraphics[scale=0.32, angle=0]{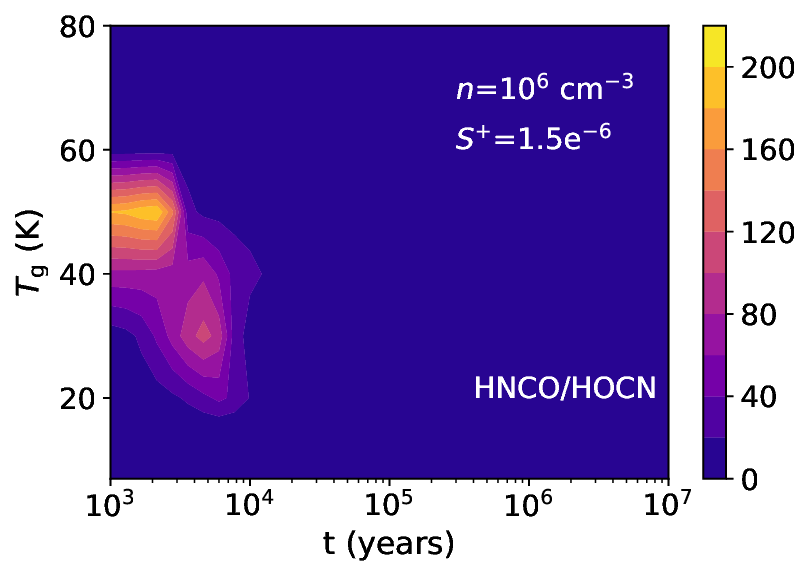} 
\hspace{-0.2cm}
\includegraphics[scale=0.32, angle=0]{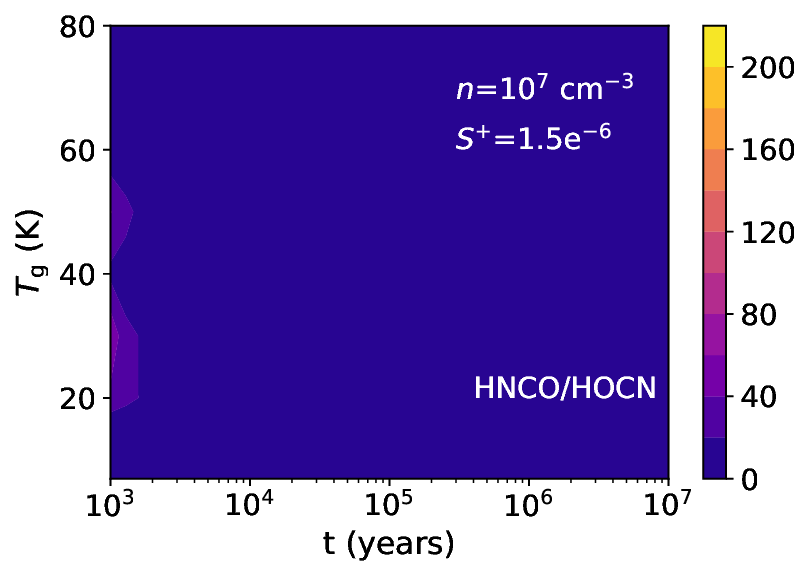}  
\hspace{-0.5cm}
\\
\caption{Evolution of HNCO/HOCN ratio (colour bar) as a function of time and temperature for a CR ionisation rate $\xi$=1.3$\times$10$^{-17}$ s$^{-1}$, an initial sulphur abundance  (relative to $n_{\mathrm{H}}$) $S$$^+$=1.5$\times$10$^{-6}$, and a hydrogen number density $n_{\mathrm{H}}$=10$^4$, 10$^5$, 10$^6$, and 10$^7$ cm$^{-3}$.
}
\label{figure:HNCO/HOCN}
\end{figure*}

\begin{figure*}
\centering
\includegraphics[scale=0.32, angle=0]{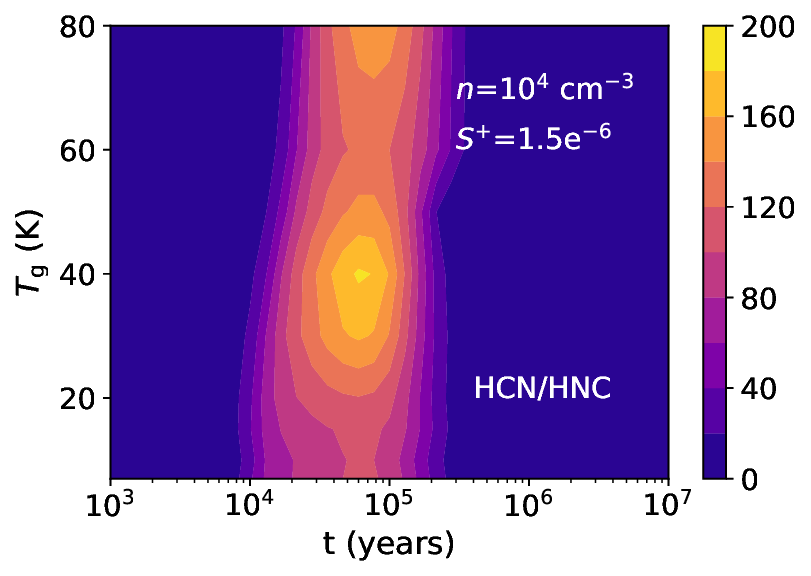}  
\hspace{-0.2cm}
\includegraphics[scale=0.32, angle=0]{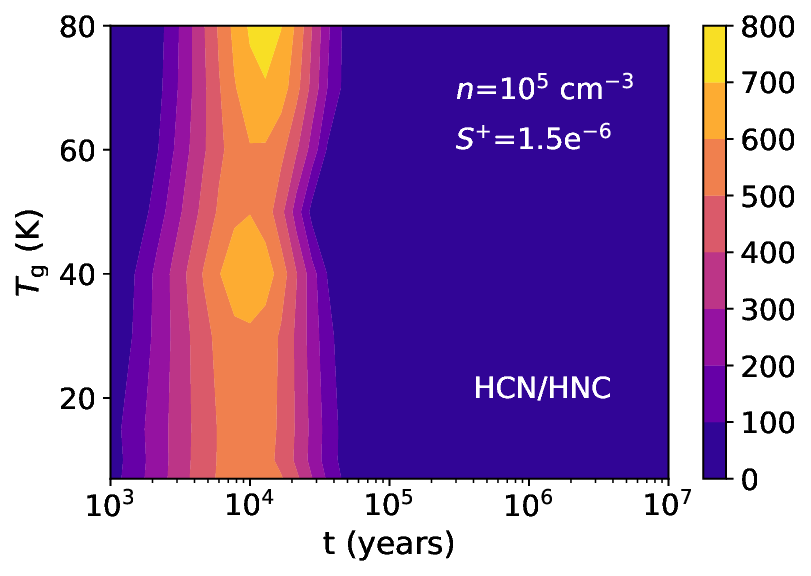}  
\hspace{-0.2cm}
\includegraphics[scale=0.32, angle=0]{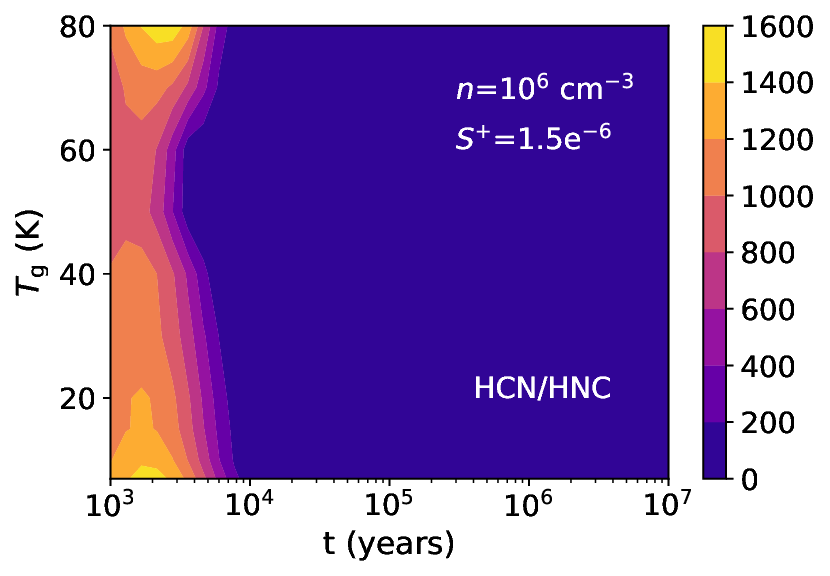} 
\hspace{-0.2cm}
\includegraphics[scale=0.32, angle=0]{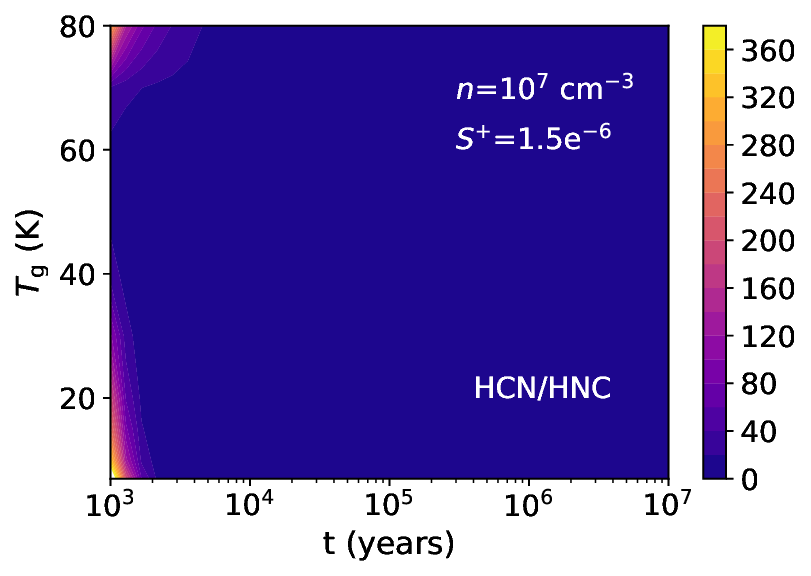}  

\hspace{-0.5cm}
\\
\caption{Evolution of HCN/HNC ratio (colour bar) as a function of time and temperature for a CR ionisation rate $\xi$=1.3$\times$10$^{-17}$ s$^{-1}$, an initial sulphur abundance  (relative to $n_{\mathrm{H}}$) $S$$^+$=1.5$\times$10$^{-6}$, and a hydrogen number density $n_{\mathrm{H}}$=10$^4$, 10$^5$, 10$^6$, and 10$^7$ cm$^{-3}$.
}
\label{figure:HCN/HNC}
\end{figure*}

\subsection{HNCS/HSCN ratio}
\label{subsection:ratio_HNCS-HSCN}

Figure \ref{figure:HNCS/HSCN} shows the evolution of the HNCS/HSCN ratio with time for different interstellar scenarios. We first observe that this ratio significantly decreases independently with temperature on the considered density for $t$$\lesssim$10$^4$ yr. In particular, the HNCS/HSCN ratio is maximal (with a value $\gtrsim$8) at temperatures lower than 30 K, and it rapidly decreases for higher temperatures and longer evolutionary times.  
In the case of a very low density ($n_{\mathrm{H}}$=10$^4$ cm$^{-3}$), we obtain the highest value with HNCS/HSCN$\sim$21, which is roughly double the maximum value obtained for higher densities. Altogether this suggests that, for $n_{\mathrm{H}}$$\leq$10$^6$ cm$^{-3}$, the HNCS/HSCN ratio may be used as a tracer of regions characterised by low temperatures. 

As previously mentioned, we observe in Fig. \ref{figure:HNCS/HSCN} that the evolutionary time also plays an important role on the HNCS/HSCN ratio, finding the highest values at early times and a significant decrease for older scenarios when $n_{\mathrm{H}}$$\leq$10$^6$ cm$^{-3}$. In particular, the highest HNCS/HSCN ratios are obtained at $t$$\lesssim$10$^4$ yr, since at these times HNCS is about one order of magnitude more abundant than HSCN (see Fig. \ref{figure:abundances}), while HSCN becomes approximately as abundant as HNCS for $t$$>$10$^4$ yr. This leads to a HNCS/HSCN ratio $\lesssim$2 for evolved stages. A high HNCS/HSCN ratio may therefore be useful for tracing very early evolutionary interstellar regions. Nevertheless, we also notice that in the case of very a high density ($n_{\mathrm{H}}$=10$^7$ cm$^{-3}$), high temperature ($T$$\sim$60 K), and $t$$>$10$^5$ yr, there is a slight increase in the HNCS/HSCN ratio, which reaches a value of up to $\sim$7. It is produced by a decrease in the HSCN abundance at these physical conditions (Fig. \ref{figure:abundances}).

\subsection{\bf{Other isomer pairs}}
\label{Other_isomers}

The fact that thiocyanic acid (HSCN) is a metastable isomer lying over 3200 K higher in energy than HNCS could suggest that it has a minor role in the interstellar chemistry compared to HNCS and that the low HNCS/HSCN ratios given in the previous section (Fig. \ref{figure:HNCS/HSCN}) for long times or high temperatures could be mainly due to a significant decrease in the HNCS abundance. However, observations of HSCN in several interstellar regions at similar abundances to those of HNCS \citep[e.g.][]{Agundez2019}, or even observations where HSCN is detected but not HNCS \citep[e.g.][]{Vastel2018}, show that there is an efficient formation of this metastable isomer in the ISM, in agreement with model results of Fig. \ref{figure:abundances}.     

This behaviour has also been detected in other isomers, such as the oxygen counterpart of HNCS, especially HNCO. The stable isomer HNCO has several metastable isomers with a singlet ground electronic state: HOCN, HCNO, and HONC. These isomers correspond to increasingly higher energies than HNCO according to quantum chemical calculations \citep[24.7, 70.7, and 84.1 kcal mol$^{-1}$, respectively, which in Kelvin scale is 12430, 35276, and 42321 K, respectively;][]{Schuurman2004}. \cite{Marcelino2009} carried out a study of HNCO and its isomers in a sample that included a variety of sources, such as quiescent prestellar cores, low-mass protostellar objects, and hot cores (some of them located in clouds of the Galactic centre, GC). HCNO was not detected towards the warm GC clouds, and only upper limits were derived. Their results showed that the HNCO/HCNO ratio is in the range between 20 and 90 in cold clouds, but it is much larger in warm clouds, suggesting that the metastable isomer HCNO is more efficiently produced in cold regions than in warm regions analogous to HSCN. Regarding HOCN (the most stable isomer after HNCO), it has been detected in several positions of the giant molecular cloud Sgr\,B2 with a HNCO/HOCN ratio $<$170 in the quiescent and extended gas at an offset position of (20$\arcsec$, 100$\arcsec$) from the hot core Sgr\,B2 (M) close to the maximum HNCO emission, and with a HNCO/HOCN ratio of up to $\sim$400 close to the hot cores Sgr\,B2 (M), (S), and (N). In the case of the cold region TMC-1, the HNCO/HOCN ratio was $\sim$100 \citep[]{Brunken2009b, Brunken2010}. These results also suggest that the metastable isomer HOCN, which seems to be as widespread as HNCO, is also more efficiently produced in cold, quiescent gas than in hot regions. Figure \ref{figure:HNCO/HOCN} shows the evolution of the HNCO/HOCN ratio with time, for different gas temperatures and densities. We observe that this ratio reaches its maximum values at very early times, in the very low-density (10$^4$ cm$^{-3}$) scenario, and for a temperature range between 30 and 50 K. In this case, the maximum abundance of HNCO is about three orders of magnitude higher than the one of HOCN. Nevertheless, the HNCO/HOCN ratio significantly decreases out of this temperature range or with the evolutionary time. For scenarios with higher density, we find a similar trend for HNCO/HOCN, but with maximum values in this ratio of up to $\sim$200. In any case, for $t$$>$5$\times$10$^4$ yr, and independently of the considered gas temperature or density, we obtain values of HNCO/HOCN$<$20.

\begin{figure*}
\centering
\includegraphics[scale=0.30, angle=0]{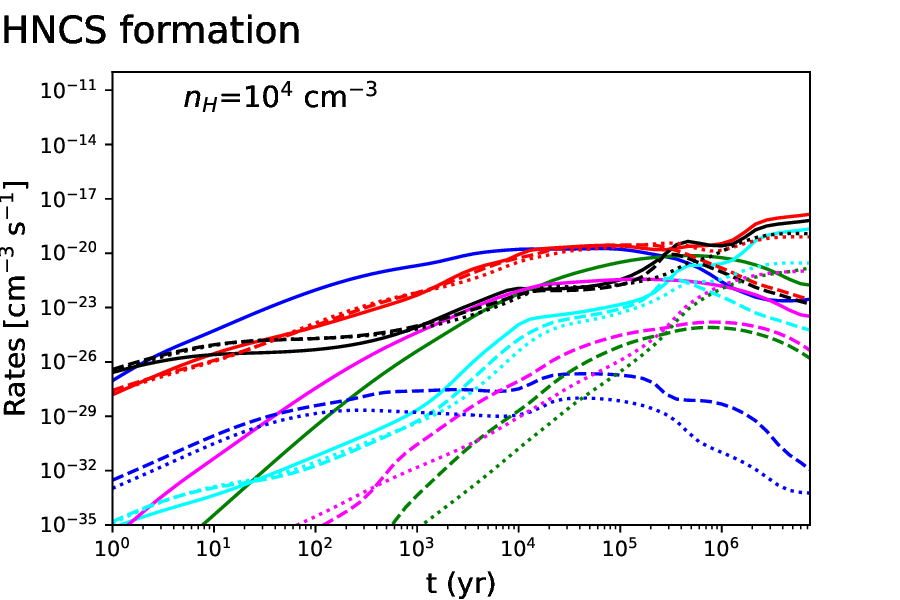}  
\hspace{-0.2cm}
\includegraphics[scale=0.30, angle=0]{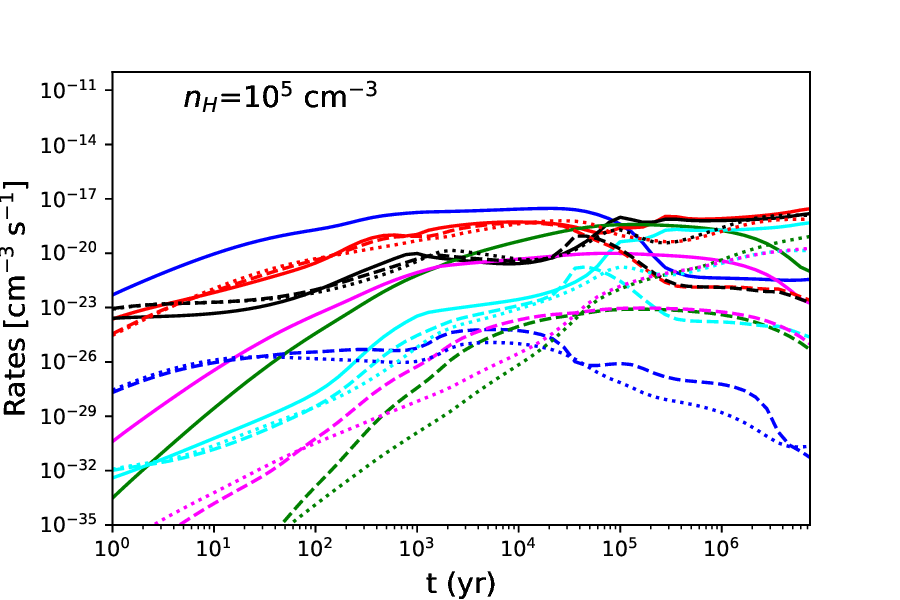}  
\hspace{-0.2cm}
\includegraphics[scale=0.30, angle=0]{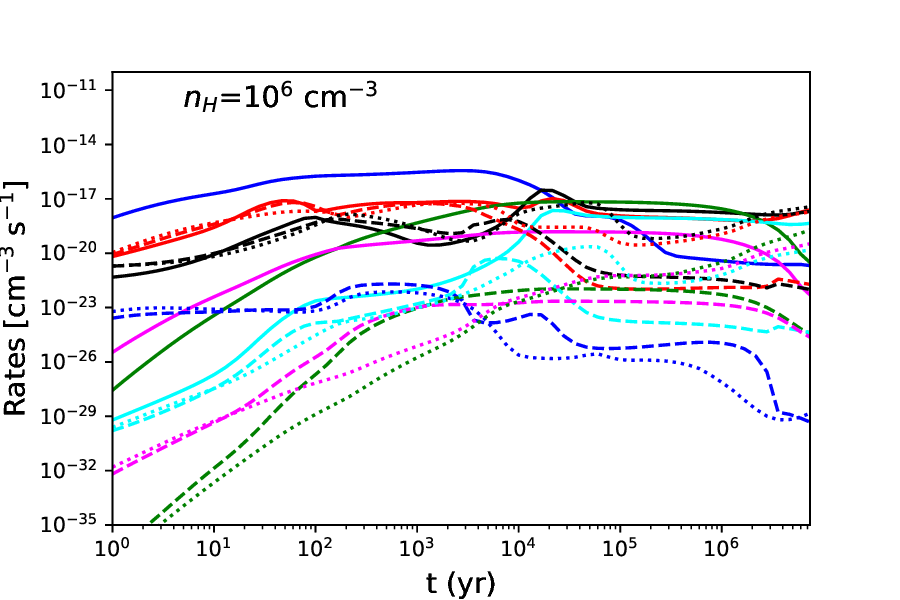} 
\hspace{-0.2cm}
\includegraphics[scale=0.30, angle=0]{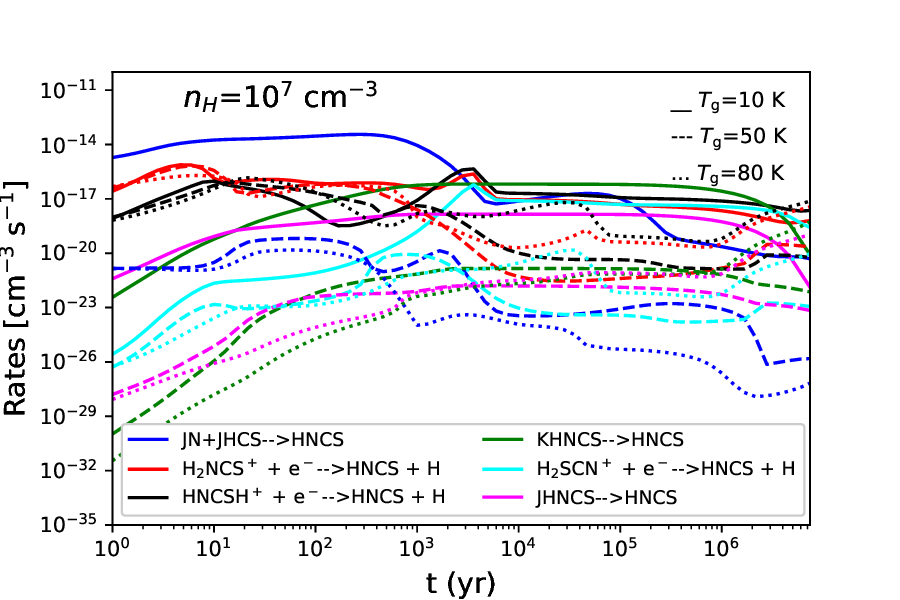}  

\includegraphics[scale=0.30, angle=0]{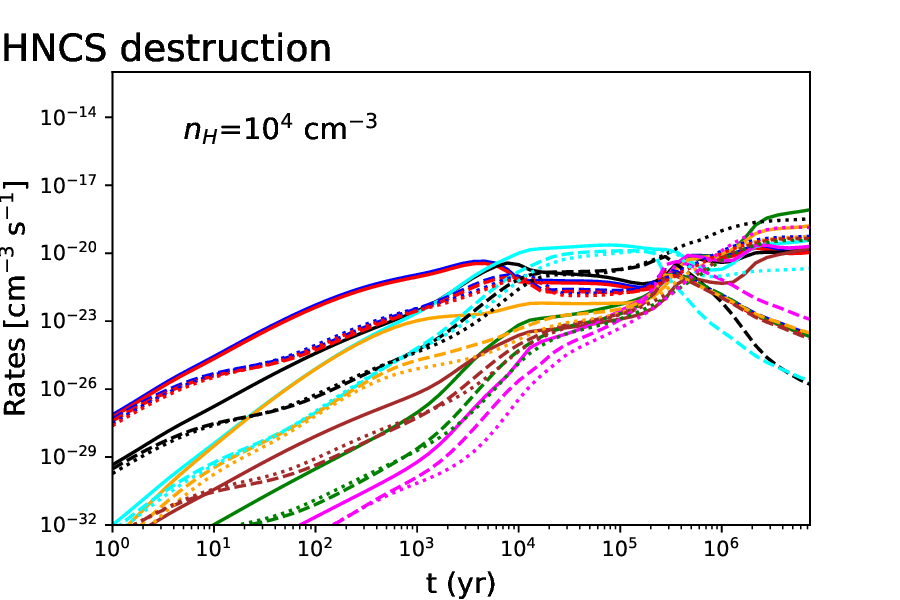}  
\hspace{-0.2cm}
\includegraphics[scale=0.30, angle=0]{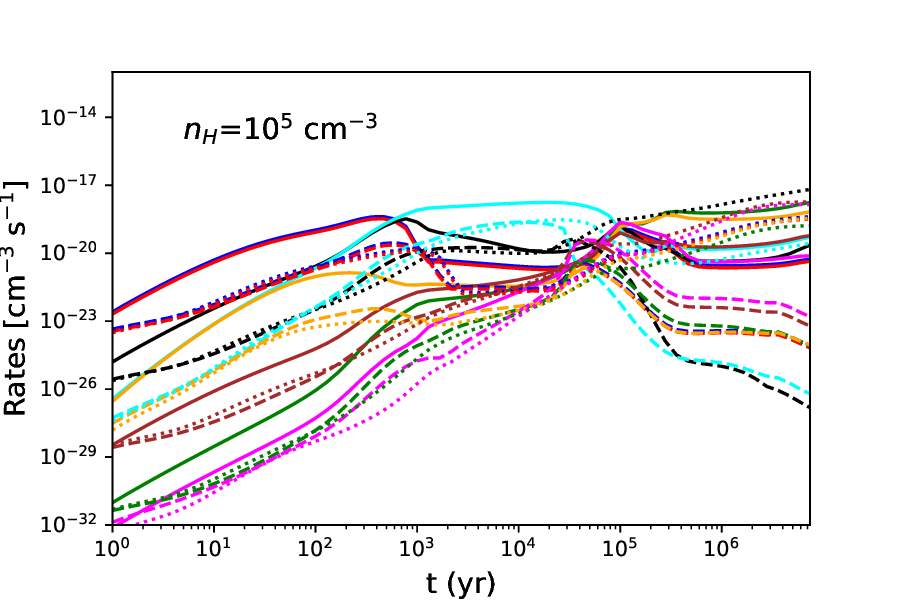}  
\hspace{-0.2cm}
\includegraphics[scale=0.30, angle=0]{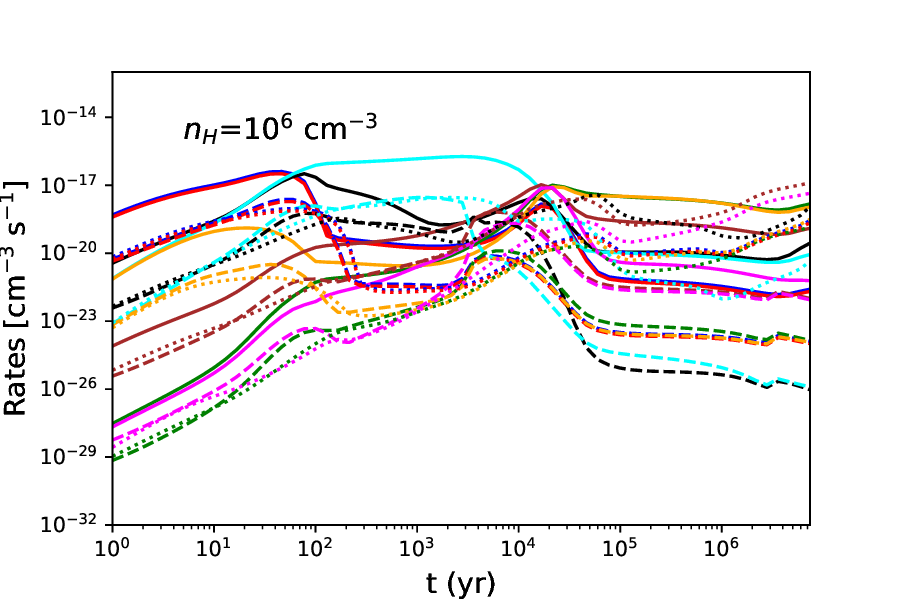}  
\hspace{-0.2cm}
\includegraphics[scale=0.30, angle=0]{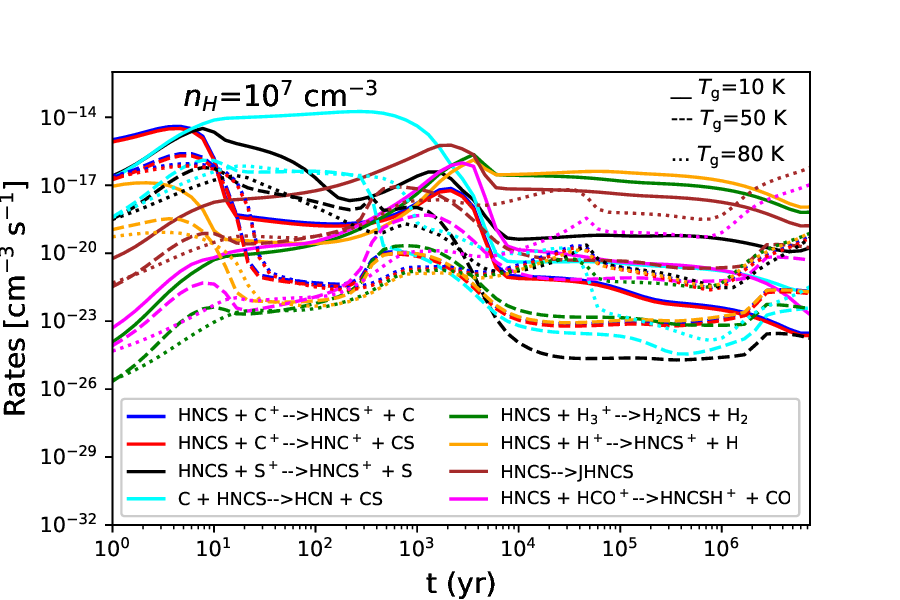}  

\hspace{-0.5cm}
\\
\caption{Main chemical reaction rates forming (top) and destroying (bottom) HNCS. JX means solid X. 
}
\label{figure:rates_HNCS}
\end{figure*}

\begin{figure*}
\centering
\includegraphics[scale=0.30, angle=0]{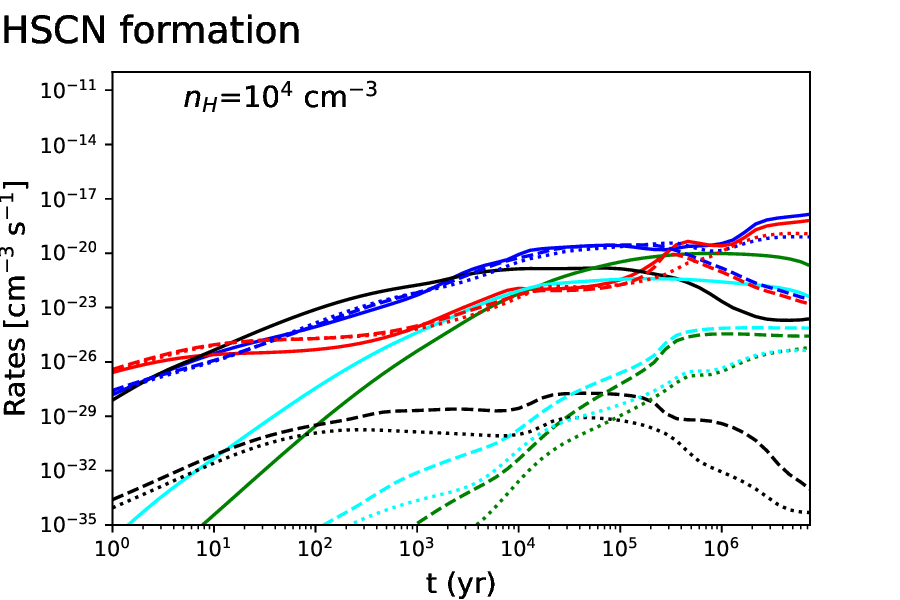}  
\hspace{-0.2cm}
\includegraphics[scale=0.30, angle=0]{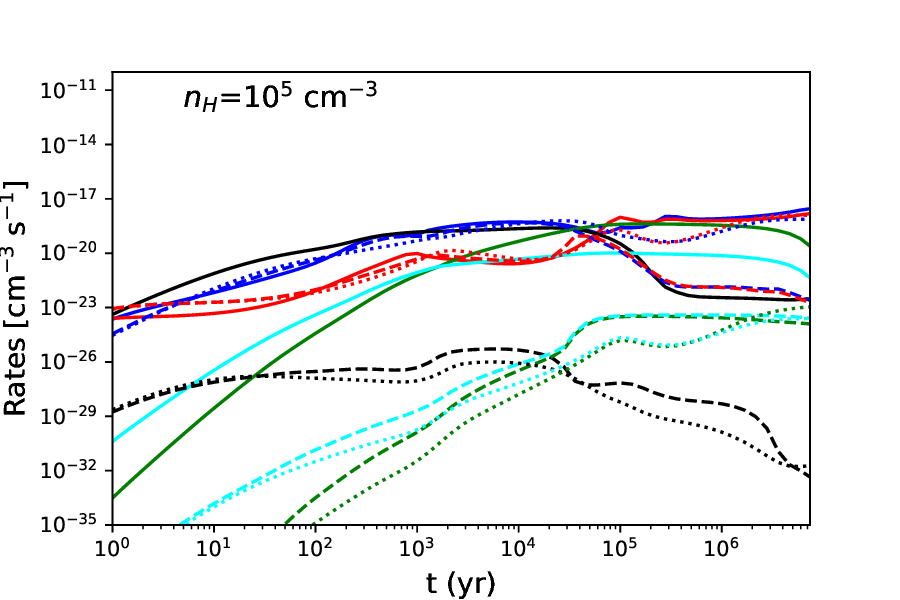}  
\hspace{-0.2cm}
\includegraphics[scale=0.30, angle=0]{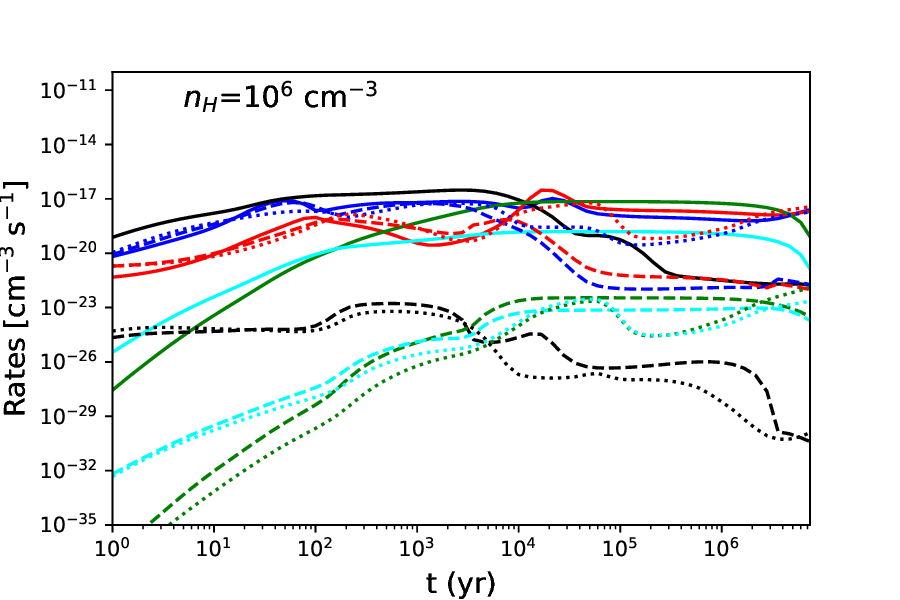} 
\hspace{-0.2cm}
\includegraphics[scale=0.30, angle=0]{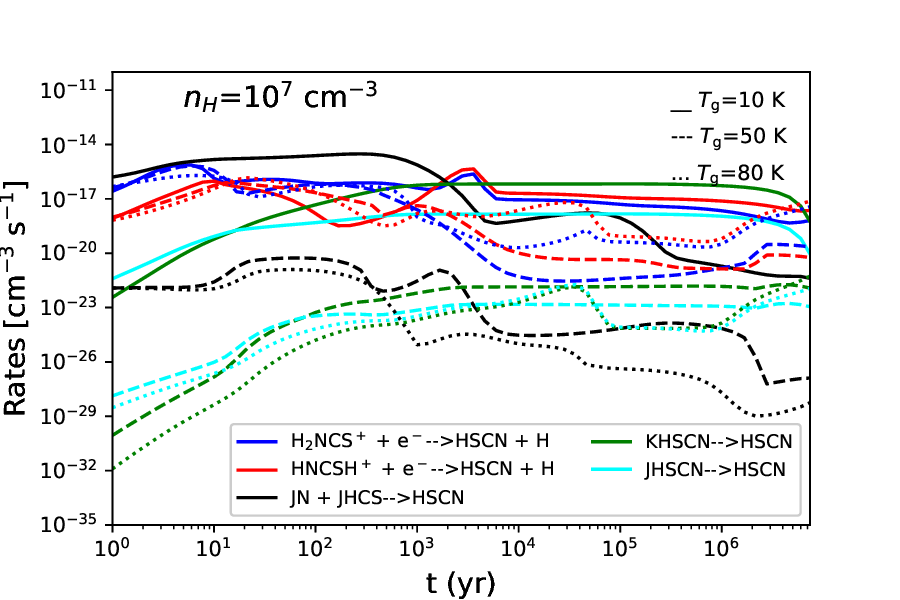}  

\includegraphics[scale=0.30, angle=0]{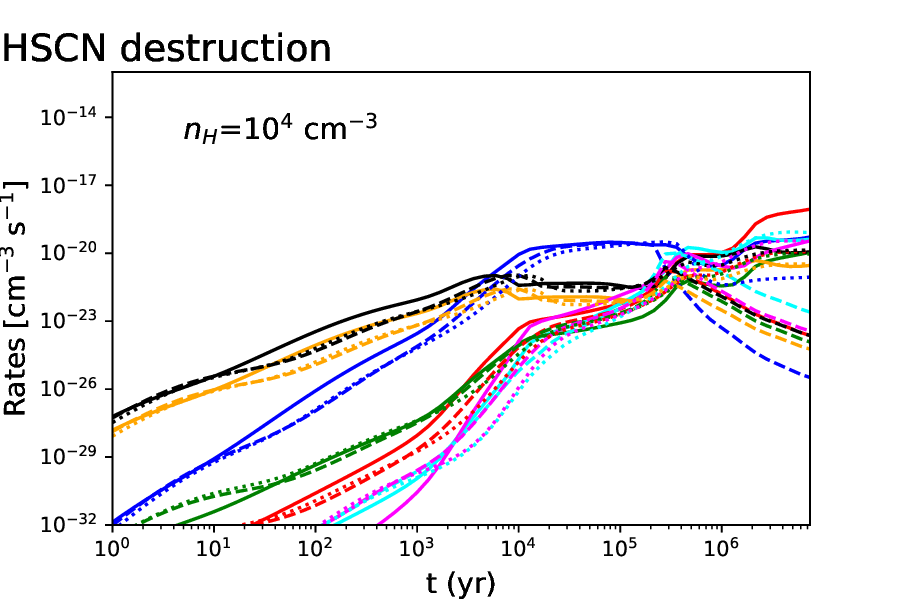}  
\hspace{-0.2cm}
\includegraphics[scale=0.30, angle=0]{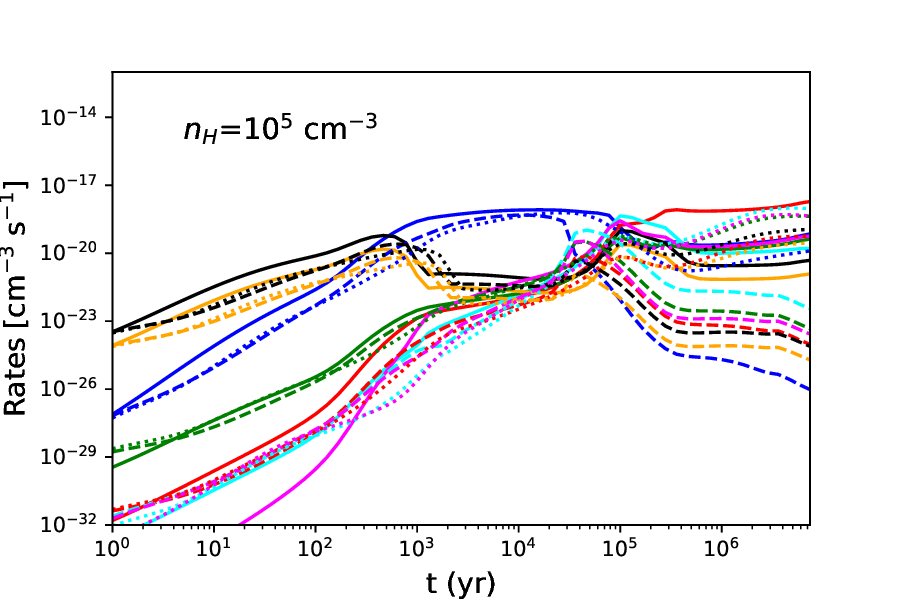}  
\hspace{-0.2cm}
\includegraphics[scale=0.30, angle=0]{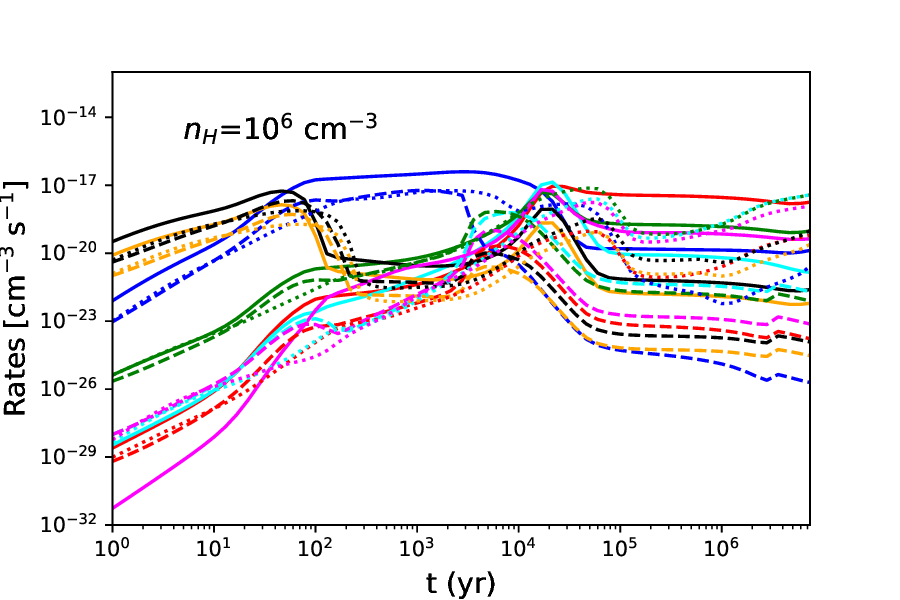}  
\hspace{-0.2cm}
\includegraphics[scale=0.30, angle=0]{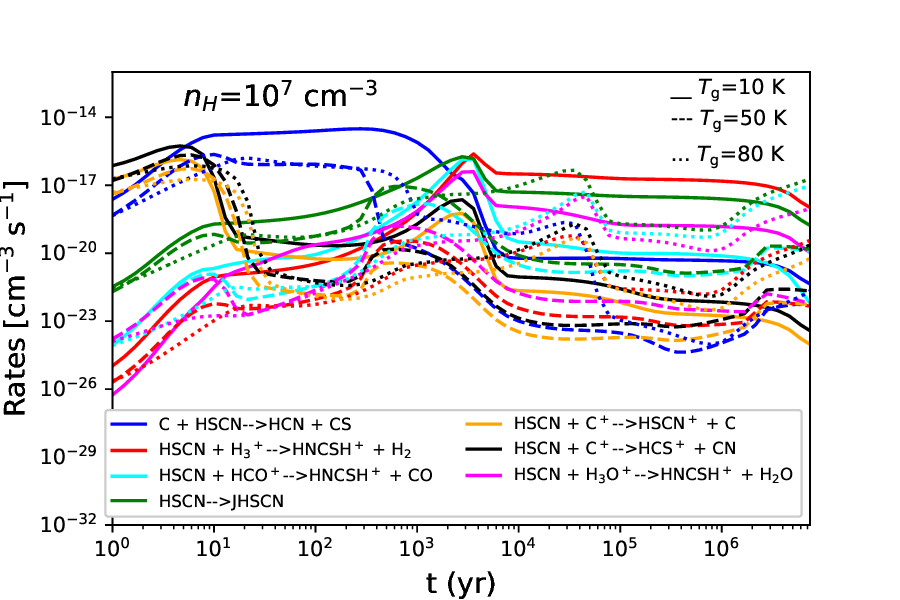}  

\hspace{-0.5cm}
\\
\caption{Main chemical reaction rates forming (top) and destroying (bottom) HSCN. JX means solid X. 
}
\label{figure:rates_HSCN}
\end{figure*}

Another isomeric pair of interest from an astrochemical point of view is HCN and HNC, were HNC is less stable than HCN by 55 kJ mol$^{-1}$; i.e., 6620 K \citep[]{Hansel1998, Baulch2005, DePrince2008}. In this case, the most stable isomer is not necessarily the most abundant in molecular clouds, especially in cold sources, since HCN and HNC also present a different behaviour depending on the kinetic temperature of the source \citep[e.g.][]{Hacar2020}. In particular, while the abundance of the metastable isomer HNC is similar to, and even higher than that of HCN in dark clouds, the HCN/HNC ratio is much larger than unity in warm molecular clouds \citep[]{Irvine1984, Churchwell1984, Schilke1992, Hirota1998, Loison2014}. The main reason for this behaviour is HNC, whose abundance decreases when the temperature increases, while the HCN abundance does not change significantly \citep[]{Marcelino2010}. \cite{Hacar2020} demonstrated, with models and observations, the strong sensitivity of the $I$(HCN)/$I$(HNC) ratio to the gas kinetic temperature and the use of this ratio as a new chemical thermometer for the molecular ISM since HCN/HNC shows increasing values at high gas temperatures. In contrast to the HNCS/HSCN ratio, which decreases with temperature (Fig. \ref{figure:HNCS/HSCN}), we also obtain the HCN/HNC trend found by \cite{Hacar2020}, as shown in Fig. \ref{figure:HCN/HNC}, although with the presence of two temperature ranges where HCN/HNC is maximised. Regarding the density effect, we find that the higher the density the larger the maximum HCN/HNC ratio, unlike for the HNCS/HSCN and HNCO/HOCN ratios. In any case, all these isomer ratios have in common the fact that they reach their maximum values at early times, i.e., $t$$<$10$^5$ yr, and then they significantly decrease.

\section{Discussion}
\label{Discussion}

\subsection{Formation and destruction mechanisms}
\label{chemical_rates}

Isomerism is a common phenomenon in interstellar chemistry \citep[]{Hollis2005}. The relative abundance of isomeric pairs of small molecules is a sensitive probe of the physical and chemical properties of astronomical sources \citep[e.g.][]{Schilke1992}. In particular, the astronomical detection of several high-lying isomers, such as HOCN, HCNO, and HSCN \citep[]{Brunken2009a, Marcelino2010, Halfen2009}, highlights the importance of non-equilibrium chemistry, putting into question our understanding of interstellar chemistry and revealing the need to include isomers of known interstellar molecules in astrochemical networks. This is especially relevant for the case of HSCN since most chemical models used in the past have poorly predicted its formation due to a lack of laboratory experiments with this isomer. In fact, up to about a decade ago, HSCN was only characterised experimentally at low spectral resolution by matrix-IR spectroscopy, forming it by UV-photolysis of HNCS in solid argon and nitrogen \citep[]{Wierzejewska2001}. Later, \cite{Brunken2009b} provided a precise determination of the rotational spectrum of HSCN by Fourier transform microwave spectroscopy between 10 and 35 GHz, as well as a millimetre-wave absorption spectroscopy between 75 and 350 GHz.

When HSCN was first detected by \cite{Halfen2009}, they deduced that it could be formed through a common ionic precursor (likely HNCSH$^+$) with HNCS, due to the similar abundances of HSCN and HNCS. A similar conclusion was reached in \cite{Adande2010} specifying that these molecules may be mainly produced in the gas phase given the energy difference of over 3200 K between the two of them. However, \cite{Cernicharo2024} recently used a gas-phase chemical model of a cold dense cloud to reproduce their observations of HNCS and HSCN in TMC-1, and the model results severely underestimate the observed abundances. To shed light on the formation and destruction of HNCS and HSCN, we studied the formation and destruction chemical rates of the main reactions related to this isomer pair, considering different astrophysical scenarios through the gas-grain chemical code Nautilus. In particular, we ran models to evaluate the evolution of the chemical rate over 10$^7$ yr for densities from 10$^4$-10$^7$ cm$^{-3}$, and three different gas temperatures, $T$$_{\mathrm{g}}$=10, 50, and 80 K. Results for HNCS and HSCN are shown in Figs. \ref{figure:rates_HNCS} and \ref{figure:rates_HSCN}, respectively. 

Regarding HNCS (top row of Fig. \ref {figure:rates_HNCS}), in a low-temperature ($T$$_{\mathrm{g}}$=10 K) scenario, the gas-grain chemistry dominates its formation  through the chemical desorption reaction JN+JHCS$\rightarrow$HNCS, where the prefix J represents the grain surface species. In particular, this surface chemical reaction dominates for $t$$<$10$^4$ yr when the density is $n$$_{\mathrm{H}}$$\geq$10$^6$ cm$^{-3}$ and for $t$$\lesssim$10$^5$ yr when the density is lower ($n$$_{\mathrm{H}}$=10$^4$-10$^5$ cm$^{-3}$). These moderate densities ($\lesssim$10$^5$ cm$^{-3}$) and low ($T$$_{\mathrm{g}}$$\sim$10 K) temperatures are the typical physical conditions found in the outer envelopes of young cold sources, such as the Class 0 objects L\,483 \citep[]{Agundez2019}, B1-b \citep[]{Daniel2013}, and B\,335 \citep[]{Shirley2011, Cabedo2023, Esplugues2023}, as well as in the cold dark core TMC-1 \citep[]{Bujarrabal1981, Pratap1997}. 
Our results indicate that the formation of HNCS in cold young objects is mainly through the surface reaction JN+JHCS$\rightarrow$HNCS unlike previous predictions \citep[e.g.][]{Halfen2009, Adande2010}. Only for $t$$>$10$^{5}$ yr, the gas-phase reactions through the ions H$_2$NCS$^+$ and HNCSH$^+$ become key to forming HNCS. 

In the cases with higher temperatures ($T$$_{\mathrm{g}}$$\geq$50 K), thermal desorption (JHNCS$\rightarrow$HNCS) barely plays a role forming HNCS. This molecule is mainly formed through gas-phase reactions, in particular, through the ion-electron dissociative recombination process H$_2$NCS$^+$+e$^-$$\rightarrow$HNCS+H, and also over long time periods by the reaction HNCSH$^+$+e$^-$$\rightarrow$HNCS+H. The ions H$_2$NCS$^+$ and HNCSH$^+$ are formed in turn through hydrogenation of the cation NCS$^+$ via H-atom transfer reactions with molecular hydrogen: NCS$^+$+H$_2$$\rightarrow$HNCS$^+$+H and HNCS$^+$+H$_2$$\rightarrow$HNCSH$^+$/H$_2$NCS$^+$+H. 

With respect to the destruction of HNCS (bottom row of Fig. \ref {figure:rates_HNCS}), for $T$$_{\mathrm{g}}$=10 K its destruction is mainly dominated by gas-phase reactions, in particular by the reaction of HNCS with carbon (ionised carbon during the earliest evolutionary stages and neutral carbon for times up to $t$$\sim$10$^5$ yr if $n$$_{\mathrm{H}}$$\leq$10$^5$ cm$^{-3}$ and up to $t$$\sim$10$^4$ yr if $n$$_{\mathrm{H}}$$\geq$10$^6$ cm$^{-3}$). Over longer time periods, hydrogen ions (both H$^{+}_{3}$ and H$^+$) are the main destroyers of HNCS. In the warm scenario ($T$$_{\mathrm{g}}$=50 K), the reaction of HNCS with the ion HCO$^+$ represents the main destruction mechanisms of HNCS at $t$$\gtrsim$10$^5$ yr and $n$$_{\mathrm{H}}$$\leq$10$^6$ cm$^{-3}$. 
For high temperatures ($T$$_{\mathrm{g}}$=80 K) and moderate densities ($n$$_{\mathrm{H}}$$\leq$10$^5$ cm$^{-3}$), carbon atoms and sulphur ions become the main destroyers of HNCS at $t$$>$10$^4$ and $t$$>$10$^5$ yr, respectively. 

Figure \ref{figure:rates_HSCN} shows the results for the formation (top panel) and destruction (bottom panel) of the metastable isomer HSCN. For a low temperature ($T$$_{\mathrm{g}}$=10 K) and moderate density ($n$$_{\mathrm{H}}$$\leq$10$^5$ cm$^{-3}$), the formation of HSCN is dominated by both gas-grain and gas-phase reactions. In particular, the chemical desorption reaction JN+JHCS$\rightarrow$HSCN, and the ion-electron recombination reaction H$_2$NCS$^+$+e$^-$$\rightarrow$HSCN+H dominate up to $t$$\sim$10$^5$ yr. Over a longer time, the ion  HNCSH$^+$ also becomes an important reactant forming HSCN. In the case at high densities ($n$$_{\mathrm{H}}$$\geq$10$^6$ cm$^{-3}$), the formation of HSCN is mainly dominated by surface reactions. 
In warm ($T$$_{\mathrm{g}}$=50 K) and hot ($T$$_{\mathrm{g}}$=80 K) scenarios (independently of the density), HSCN is mostly formed through the ions H$_2$NCS$^+$ and HNCSH$^+$.

Regarding the destruction of HSCN, in cold regions ($T$$_{\mathrm{g}}$=10 K), it is mainly destroyed through its reaction with ionised carbon (early times), then by reacting with neutral carbon, and also by reacting with ions of hydrogen (H$^{+}_{3}$, for long times). We find this behaviour independently of the density. When $T$$_{\mathrm{g}}$ increases ($T$$_{\mathrm{g}}$$\geq$50 K), the HSCN destruction caused by the reaction with HCO$^+$ becomes significant at late times ($t$$>$10$^5$ yr).

\subsection{Observations of HNCS and HSCN in the ISM: The case of B1-a}
\label{Observations}

We have carried out observations of B1-a (03$^{\mathrm{h}}$:33$^{\mathrm{m}}$:16.67$^{\mathrm{s}}$,  31$^{\mathrm{o}}$:07$^{\mathrm{'}}$:55.1$^{\mathrm{''}}$) with the Yebes 40-m telescope located at Guadalajara (Spain). In these observations, we used a receiver that consists of two cold, high electron-mobility transistor amplifiers covering the 72-90 GHz W band with horizontal and vertical polarisations. The backends are 2$\times$8$\times$2.5 GHz fast Fourier transform spectrometers with a spectral resolution of 38 kHz, providing full coverage of the W band in both polarisations. The observational mode was position-switching with (-400$\arcsec$, 0$\arcsec$) as the reference position. The main beam efficiency varies from 0.3 at 72.5 GHz to 0.21 at 88.5 GHz.

The Solar-type protostar (Class I) B1-a is located in Perseus \citep[]{Boogert2008, Graninger2016} and stands out as particularly line rich. This source is, in addition, in the vicinity of another YSO, B1-b, which is a known host of complex molecules \citep[]{Oberg2014, Marcelino2018}. B1-a also has an outflow from the protostellar core, as revealed by maps of the shock tracer SiO \citep[]{Bergner2019}.

\begin{figure}
\centering
\includegraphics[scale=0.6, angle=0]{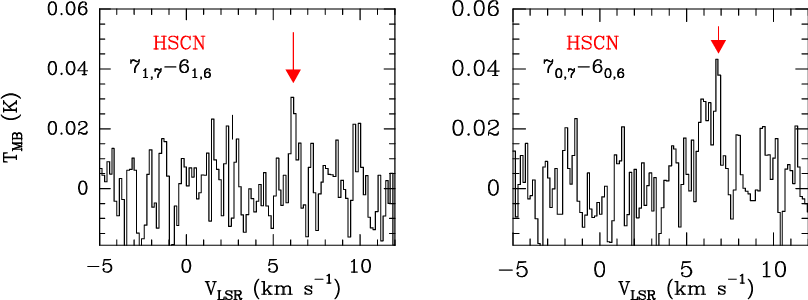} 
\hspace{-0.5cm}
\\
\caption{Observed lines of HSCN in B1-a.}
\label{figure:observed_HSCN}
\end{figure}

\begin{table*}
\centering
\caption{Line parameters obtained from Gaussian fits of HSCN lines in B1-a.}
\begin{tabular}{lllllllll}
\hline 
\hline
Species    & Transition   & Frequency    & $E$$_{\mathrm{l}}$ & $A$$_{\mathrm{ul}}$ &  v$_{\mathrm{LSR}}$ & $\Delta$v    & $T$$_{\mathrm{MB}}$  & $\int$$T$$_{\mathrm{MB}}$dv  \\
           &              & (MHz)        &  (K)               & (s$^{-1}$)            & (km s$^{-1}$)        & (km s$^{-1}$)  & (K)                  & (K km s$^{-1}$)                \\
\hline
\hline 
HSCN       & 7$_{1}$$_{,}$$_{7}$-6$_{1}$$_{,}$$_{6}$  &  79863.71  & 25.1 & 3.24e-5 & 6.09$\pm$0.06         & 0.4$\pm$0.1   & 0.030$\pm$0.008        & 0.012$\pm$0.004 \\
HSCN       & 7$_{0}$$_{,}$$_{7}$-6$_{0}$$_{,}$$_{6}$  &  80283.19  & 11.6 & 3.36e-5 & 6.62$\pm$0.05          & 0.5$\pm$0.1   & 0.043$\pm$0.010        & 0.021$\pm$0.003 \\
\hline
\end{tabular}
\label{table:HSCN_line_parameters}
\end{table*}

\begin{table*}
\centering
\caption{Observed column densities, $N$ (cm$^{-2}$), of HNCS and HSCN in different types of sources.}
\begin{tabular}{lllllll}
\hline 
\hline
Source  & Source type                & $T$$_{\mathrm{K}}$ (K)   & $N$$_{\mathrm{HNCS}}$  (cm$^{-2}$)                &  $N$$_{\mathrm{HSCN}}$ (cm$^{-2}$)             & $N$$_{\mathrm{HNCS}}$/$N$$_{\mathrm{HSCN}}$ & References \\
\hline
\hline 
TMC-1   & Starless core          &       & (8.3$\pm$4.4)$\times$10$^{10}$     & (6.4$\pm$2.1)$\times$10$^{10}$   & 1.3$\pm$0.7         & \cite{Adande2010}   \\
        &               &   9.1     &   (3.2$\pm$0.1)$\times$10$^{11}$     &  (8.3$\pm$0.4)$\times$10$^{11}$ &  0.39$\pm$0.25      & \cite{Cernicharo2024}   \\
\hline
B\,213 C16 & Starless core        &    $\lesssim$10    & -           & (5.7$\pm$2.8)$\times$10$^{10}$  & -         & Moral-Almansa et al. in prep.   \\ 
\hline
L\,1544 & Pre-stellar core      & 7.5         & -                                  & (6.0$\pm$0.2)$\times$10$^{10}$  & -                   & \cite{Vastel2018}, \\
        &          &         &                                   &   &                    & \cite{Bianchi2023} \\ 
\hline
L\,483  & Class 0               & 9.5        & (2.1$\pm$1.8)$\times$10$^{11}$     & (1.1$\pm$0.4)$\times$10$^{11}$  & 1.9$\pm$0.5         & \cite{Anglada1997}, \\
        &                       &            &      &   &         & \cite{Agundez2019} \\ 
\hline
IRAS\,4A  & Class 0             & 40           & $<$7.1$\times$10$^{11}$     & (1.5$\pm$0.7)$\times$10$^{11}$  & $<$4.7         & Fernández-Ruiz et al. in prep., \\
        &                       &          &  &  & & \cite{Koumpia2016}, \\ 
                &                       &          &  &  & & \cite{Quitian2024} \\ 
\hline
B\,335  & Class 0               & 15       & $<$4.7$\times$10$^{12}$  & (1.0$\pm$0.5)$\times$10$^{12}$  & $<$4.7           & \cite{Shirley2011}, \\ 
        &                       &          &  &  &          & \cite{Esplugues2023} \\ 
\hline
B1-a    & Class I               & 15        & $<$1.45$\times$10$^{12}$  & (1.1$\pm$0.6)$\times$10$^{12}$  & $<$1.3              & This work         \\
\hline
Sgr\,B2 & High-mass  & $\gtrsim$150 & (0.8-4.7)$\times$10$^{13}$        &  (0.2-1.3)$\times$10$^{13}$      & 2.2-7.0                 & \cite{Bonfand2019}, \\
        & star-forming region  &  &        &       &           & \cite{Adande2010} \\

\hline
\end{tabular}
\label{table:N_sources}
\end{table*}

We detected two transitions of HSCN with an rms $>$3$\sigma$ (Fig. \ref{figure:observed_HSCN}), but none from HNCS. The detected HSCN transitions span an energy range of $E$$_{\mathrm{up}}$=15-29 K. These two lines are observed in emission. We first fitted the observed lines with Gaussian profiles using the CLASS software to derive the radial velocity ($v$$_{\mathrm{LSR}}$), the line width, and the intensity for each line. These results are shown in Table \ref{table:HSCN_line_parameters}. The detected HSCN emission lines show narrow line profiles ($\Delta$v$\sim$0.37-0.45 km s$^{-1}$), suggesting that the HSCN arises from a quiescent region and not from a shocked region. 
This allows us to fit the line profiles using a single velocity component. 

In order to calculate the HSCN column density, we used the LVG code MADEX \citep[]{Cernicharo2012}, which includes beam dilution of each line depending on the different beam size at each frequency. The code fits the emission-line profiles assuming uniform physical conditions (kinetic temperature, density, line width, source size). For the line width, we assumed a value of 0.45 km s$^{-1}$ derived from the Gaussian fits (Table \ref{table:HSCN_line_parameters}). Regarding the source size, we assumed a size of 12$\arcsec$ corresponding to $\sim$3600 au at the distance to Perseus \citep[]{Ortiz2018} since it would include the envelope emission of B1-a according to previous studies \citep[]{Bergner2019}. In the chemical study carried out by these authors in B1-a, a gas temperature of $\sim$10 K was also obtained. We therefore ran models for two temperatures around this value ($T$$_{\mathrm{K}}$=8 K and 15 K). Regarding the density, we also ran models considering two different density values, 10$^4$ cm$^{-3}$ and 10$^5$ cm$^{-3}$, since we observe emission from the envelope of B1-a given the large beam size of the used telescope used (HPBW$\sim$22.2$\arcsec$ and 22.0$\arcsec$ for the frequencies 79863 MHz and 80283 MHz, respectively), and, in these regions, the density is usually not greater than 10$^5$ cm$^{-3}$. The best fits obtained for each group of input parameters are shown in Figs. \ref{figure:HSCN_8K_n10to4}-\ref{figure:HSCN_15K_n10to5}. From these Figures, we deduced that the model that best reproduces the detected HSCN emission lines is a model with a gas temperature of 15 K and a density of 10$^4$ cm$^{-3}$ (Fig. \ref{figure:HSCN_15K_n10to4}). This model leads to a HSCN column density of $N$$_{\mathrm{HSCN}}$=1.1$\times$10$^{12}$ cm$^{-2}$. We also used MADEX to derive the upper limit for the HNCS column density considering the same parameters used to obtain $N$$_{\mathrm{HSCN}}$. This process consists of varying the HNCS column density until the model fit reaches the observed intensity peak of any of the HNCS observed lines. We do not allow the model fit to be greater than any observed line. Following this procedure, we derive $N$$_{\mathrm{HNCS}}$$<$1.45$\times$10$^{12}$ cm$^{-2}$, which implies a HNCS/HSCN ratio $<$1.3 for the Solar-type protostar B1-a.

Classified as a Class I source, B1-a has an evolutionary age of $\sim$10$^5$ yr \citep[]{Bianchi2019}. As also previously mentioned, HSCN observations are better reproduced with MADEX when considering a temperature of 15 K and a density of 10$^4$ cm$^{-3}$. Taking these numbers into account to compare with theoretical results, chemical models of Fig. \ref{figure:HNCS/HSCN} (left panel) show that the HNCS/HSCN is $<$3 for those physical conditions, in agreement with the observational results (HNCS/HSCN $<$1.3, Table \ref{table:N_sources}). This low ratio highlights the significant presence of HSCN in this type of source, in spite of it lying 3200 K higher in energy than the stable isomer HNCS (as also shown in Fig. \ref{figure:abundances}). In fact, Fig. \ref{figure:abundances} (left panels) shows that HSCN is roughly as abundant as HNCS for $t$$\gtrsim$3$\times$10$^5$ yr, with an abundance $\gtrsim$10$^{-12}$. However, in spite of model results indicating that HNCS and HSCN have similar abundance, we only detect HSCN in B1-a. The detection of emission from one metastable isomer (HSCN) and not from the stable one (HNCS) was also the case for \cite{Vastel2018} in the L\,1544 pre-stellar core. In this instance, a column density of $N$$_{\mathrm{HSCN}}$=(6.0$\pm$0.2)$\times$10$^{10}$ cm$^{-2}$ was derived, which is a factor of $\sim$20 below our value in B1-a. In the Taurus region, HSCN emission ($N$=(5.7$\pm$2.8)$\times$10$^{10}$ cm$^{-2}$) has been detected (Moral-Almansa et al. in prep.) in the starless core C16, which is located in the B\,213 filament \citep[]{Hacar2013, Rodriguez-Baras2021, Esplugues2022, Fuente2023}, while HNCS emission was not observed. In another starless core (TMC-1), however, \cite{Adande2010} detected both isomers with column densities similar to those found for HSCN in L\,1544 (with HNCS/HSCN ratio of 1.3), while the column densities deduced by \cite{Cernicharo2024} in TMC-1 for HNCS and HSCN are a factor of $\sim$4 and $\sim$13, respectively, which is higher than those found by \cite{Adande2010}. In the Class 0 object B\,335, \cite{Esplugues2023} also detected HSCN emission with $N$$_{\mathrm{HSCN}}$=(1.0$\pm$0.5)$\times$10$^{12}$ cm$^{-2}$, while HNCS was only tentatively detected with $N$$_{\mathrm{HNCS}}$$<$4.7$\times$10$^{12}$ cm$^{-2}$, leading to a HNCS/HSCN ratio of $<$4.7. In IRAS\,4A (another Class 0 object); thus, only HSCN was detected ($N$$_{\mathrm{HSCN}}$=(1.5$\pm$0.7)$\times$10$^{11}$ cm$^{-2}$, Fernández-Ruiz et al. in prep.), not HNCS ($N$$_{\mathrm{HNCS}}$$<$7.1$\times$10$^{11}$ cm$^{-2}$). Both molecules were, however, clearly detected in the Class 0 object L\,483, with a HNCS/HSCN ratio=1.9 \citep[]{Agundez2019}. The fact that HNCS is detected in some sources but not in others of the same type suggests that this molecule might be quite sensitive to the specific environmental conditions of each source, rather than an overproduction (or missing destruction mechanism) of HNCS in the chemical code occurring. If we consider interstellar sources characterised by higher temperatures, we only find detections of HNCS and/or HSCN in Sgr\,B2 \citep[]{Halfen2009, Adande2010}. In this case, the column densities for both isomers are significantly higher than those found in the colder sources (starless cores, Class 0, and Class I), as shown in  Table \ref{table:N_sources}, with values ranging from (0.8-4.7)$\times$10$^{13}$ cm$^{-2}$ and (0.2-1.3)$\times$10$^{13}$ cm$^{-2}$ for HNCS and HSCN, respectively. The HNCS/HSCN ratios in Sgr\,B2 have a range of 2.2–7.0, with the largest values near the 2N position, where there are no hot cores and the gas should be relatively quiescent \citep[]{Adande2010}.

\begin{figure}
\centering
\includegraphics[scale=0.6, angle=0]{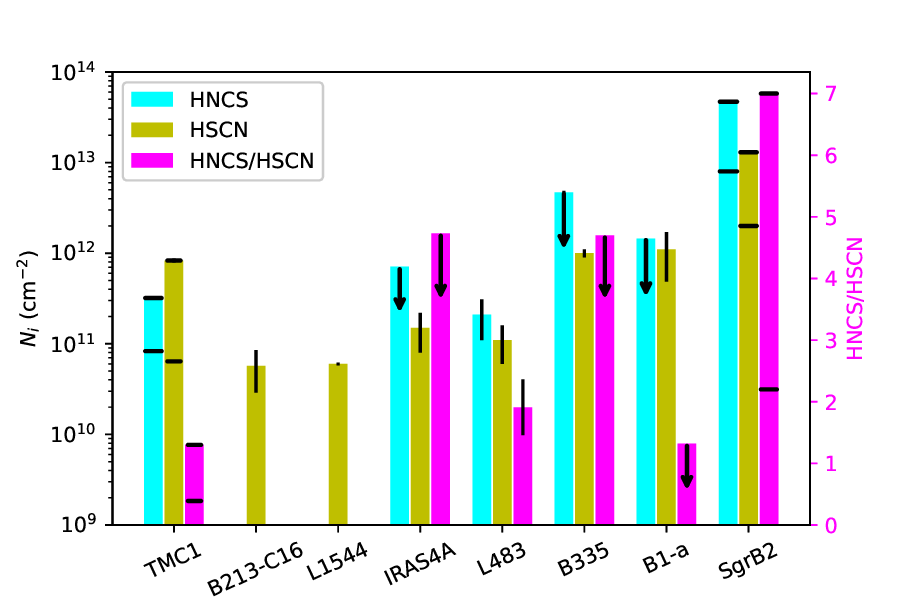} 

\hspace{-0.5cm}
\\
\caption{HNCS and HSCN column densities (left Y-axis) and HNCS/HSCN ratio (right Y-axis) for different objects: starless/pre-stellar cores (TMC-1, L\,1544, B\,213-C16), Class 0 (IRAS\,4A, L\,483, B\,335), Class I (B1-a), and a high-mass star forming region (Sgr\,B2). Black arrows indicate upper limits (B\,335 and B1-a). For TMC-1 and Sgr\,B2, the observed column density ranges are indicated with horizontal lines. References to these data are included in Table \ref{table:N_sources}.
}
\label{figure:comparison_sources}
\end{figure}

Figure \ref{figure:comparison_sources} shows values of $N$$_{\mathrm{HSCN}}$ in B\,335 (Class 0) and B1-a (Class I) that are about one order of magnitude higher than those found in some younger objects, such as the starless/pre-stellar cores B\,213-C16 and L\,1544. It could suggest a correlation between $N$$_{\mathrm{HSCN}}$ and the source evolutionary stage. However, the values of $N$$_{\mathrm{HSCN}}$ (and also of $N$$_{\mathrm{HNCS}}$) found in L\,483 and IRAS\,4A (both Class 0) are significantly lower than the ones found in B\,335 (Class 0). This is also the case of the $N$$_{\mathrm{HSCN}}$ values found in the starless/pre-stellar cores B\,213-C16 and L\,1544 with respect to the one found in TMC-1 by \cite{Cernicharo2024}. These column density differences in objects with the same evolutionary stage reinforce the previously mentioned conclusion that environmental effects might have a key influence on their values.   
Regarding the HNCS/HSCN ratio, we do not find a clear observational trend among the analysed sources.
Nevertheless, given the small source sample (with a very limited temperature range) in which these isomers have been detected in the ISM up to date, we would need to increase the number of different regions, especially including observations of warmer regions (such as hot cores and PDRs). This would allow us to derive a stronger conclusion on the presence of HNCS and HSCN in each type of source and the possible dependence of its ratio on the temperature as suggested by the theoretical results.

\section{Summary and conclusions}
\label{section:summary}

We carried out a comprehensive theoretical study of the sulphur isomer pair HNCS and HSCN through the time-dependent gas-grain chemical code Nautilus. This allowed us to analyse the influence of several physical parameters (density, gas temperature, time) on their abundances, as well as on the HNCS/HSCN ratio. The results show that low-density interstellar environments promote the formation of the metastable isomer HSCN, in addition to maintaining the highest abundances of both species, HNCS and HSCN, much longer once their maximum values have been reached. Regarding the gas temperature, the models show that the highest abundances are mainly reached for gas temperatures $\lesssim$40 K, or even at temperatures $\lesssim$20 K for very high-density regions. With Nautilus, we also studied the HNCS/HSCN abundance ratio considering different scenarios. We find that, for $n_{\mathrm{H}}$$\leq$10$^6$ cm$^{-3}$, this ratio significantly decreases when the gas temperature and/or the evolutionary time increases, suggesting that the HNCS/HSCN ratio may be useful as a low-temperature tracer.    

Unlike previous studies which indicated that the formation of HNCS and HSCN is driven by gas-phase ion–molecule chemistry, our theoretical analysis has revealed that in cold environments ($\sim$10 K) the formation of HNCS and HSCN is mainly determined by grain-surface chemistry (in particular, by chemical desorption through the reaction between JN and JHCS) for the first $\sim$10$^5$ yrs and $\sim$10$^4$ yrs, respectively. While for longer evolutionary times or higher gas temperatures, gas-phase chemistry dominates the formation of this sulphur isomer pair through ion-electron recombination reactions (mainly through the ions H$_2$NCS$^+$ and HNCSH$^+$). 

In this paper, we also present the observational detection of the metastable isomer HSCN in the Solar-type protostar (Class I) B1-a using the Yebes 40-m telescope. In spite of the stable isomer HNCS lying 3200 K (about 6 kcal mol$^{-1}$) lower in energy than HSCN, we do not detect HNCS in this source; thus, only an upper limit on its column density has been provided. A comparison of the presence of HNCS and HSCN in different types of regions makes evident the important influence of the environmental effects on their column densities.   
Regarding the observed HNCS/HSCN ratio, given that the available source sample where both isomers were detected is mostly formed by cold regions (starless/pre-stellar cores and Class 0 objects), the temperature range is too limited to derive strong conclusions about the behaviour of this ratio with temperature. This highlights, therefore, the need to increase the source sample by including high-temperature objects, such as hot cores and PDRs, in order to contrast the theoretical results with observations and to properly analyse the role of the sulphur isomer ratio HNCS/HSCN as a low-temperature tracer.

\begin{acknowledgements}

We thank the Spanish MICINN for funding support from the project PID2022-137980NB-I00. This project has also received funding from the European Research Council (ERC) under the European Union’s Horizon Europe research and innovation programme ERC-AdG-2022 (GA No. 101096293). M.N.D. acknowledges the Holcim Foundation Stipend. This project has been carried out with observations from the 40-m radio telescope of the National Geographic Institute of Spain (IGN) at Yebes Observatory. 
  
\end{acknowledgements}

\bibliographystyle{aa}
\bibliography{biblio}

\begin{appendix}



\section{Additional figures}
\label{tables}

\newpage

\begin{figure}
\centering
\includegraphics[scale=0.6, angle=0]{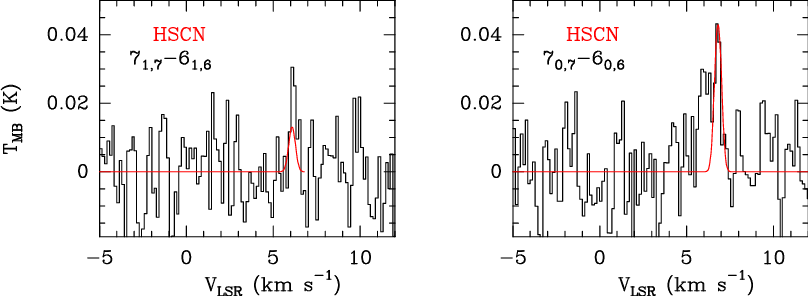} 

\hspace{-0.5cm}
\\
\caption{Observed lines of HSCN (black histogram) and best-fit LVG model results (red) obtained assuming $T$$_{\mathrm{kin}}$=8 K and $n$=10$^4$ cm$^{-3}$.
}
\label{figure:HSCN_8K_n10to4}
\end{figure}

\begin{figure}
\centering
\includegraphics[scale=0.6, angle=0]{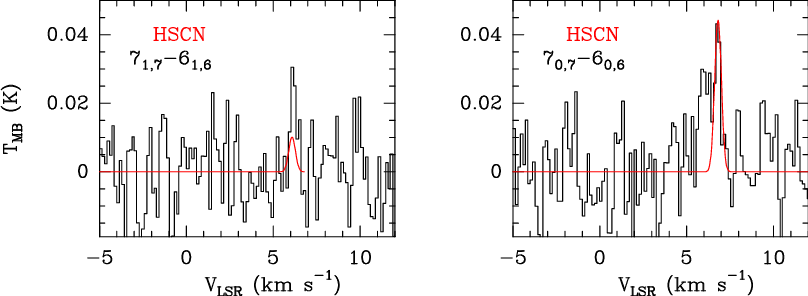} 

\hspace{-0.5cm}
\\
\caption{Observed lines of HSCN (black histogram) and best-fit LVG model results (red) obtained assuming $T$$_{\mathrm{kin}}$=8 K and $n$=10$^5$ cm$^{-3}$.
}
\label{figure:HSCN_8K_n10to5}
\end{figure}

\begin{figure}
\centering
\includegraphics[scale=0.6, angle=0]{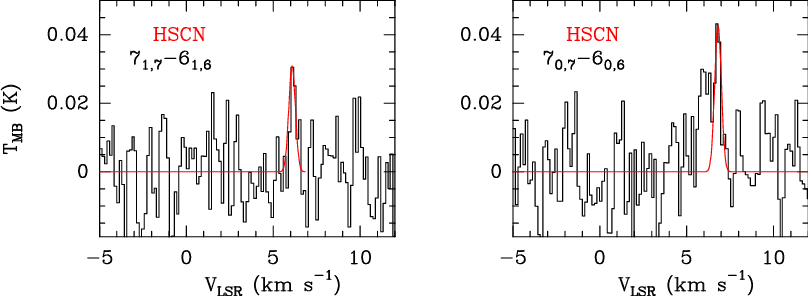} 

\hspace{-0.5cm}
\\
\caption{Observed lines of HSCN (black histogram) and best-fit LVG model results (red) obtained assuming $T$$_{\mathrm{kin}}$=15 K and $n$=10$^4$ cm$^{-3}$.
}
\label{figure:HSCN_15K_n10to4}
\end{figure}

\begin{figure}
\centering
\includegraphics[scale=0.6, angle=0]{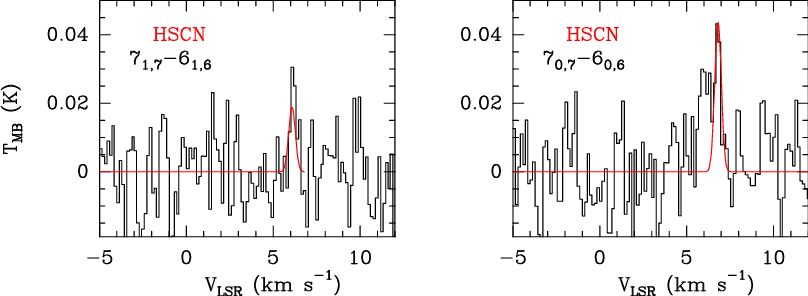} 

\hspace{-0.5cm}
\\
\caption{Observed lines of HSCN (black histogram) and best-fit LVG model results (red) obtained assuming $T$$_{\mathrm{kin}}$=15 K and $n$=10$^5$ cm$^{-3}$.
}
\label{figure:HSCN_15K_n10to5}
\end{figure}


\end{appendix}

\end{document}